\documentclass[twocolumn]{aastex631}
\newcommand{\ch}[1]{{$\rm #1$}}

\newcommand{\hl}[1]{{#1}}
\newcommand{\hll}[1]{{#1}}
\usepackage[most]{tcolorbox}
\usepackage{ragged2e}
\usepackage{float}
\usepackage{rotating}
\usepackage{multirow}
\usepackage{placeins}
\usepackage{ulem}
\newtcbox{\myboxi}[1][]{nobeforeafter,tcbox raise base,colframe=green!50!black,colback=green!50!black,height=8pt,valign=center,raster valign=center,
  box align=base,sharp corners,top=0pt,bottom=0pt,left=0pt,right=2pt,
  boxrule=0pt,boxsep=2.5pt,before upper=\strut,#1}
\newcommand{\mybox}[2][1.1ex]{\raisebox{#1}{\myboxi{#2}}}
        
\newtcbox{\myboxxi}[1][]{nobeforeafter,tcbox raise base,colframe=black!50!white,colback=black!10!white,height=8pt,valign=center,raster valign=center,
  box align=base,sharp corners,top=0pt,bottom=0pt,left=0pt,right=2pt,
  boxrule=0pt,boxsep=2.5pt,before upper=\strut,#1}
\newcommand{\myboxx}[2][1.1ex]{\raisebox{#1}{\myboxxi{#2}}}
        
\newtcbox{\myboxxxi}[1][]{nobeforeafter,tcbox raise base,colframe=yellow!85!black!,colback=yellow!85!black!,height=8pt,valign=center,raster valign=center,
  box align=base,sharp corners,top=0pt,bottom=0pt,left=0pt,right=2pt,
  boxrule=0pt,boxsep=2.5pt,before upper=\strut,#1}
\newcommand{\myboxxx}[2][1.1ex]{\raisebox{#1}{\myboxxxi{#2}}}
\usepackage{array}
\newcolumntype{P}[1]{>{\centering\arraybackslash}p{#1}}
\usepackage{graphicx}
\usepackage{orcidlink}
\usepackage{txfonts}

\emergencystretch=\maxdimen
\begin{document}

\title{MINDS: The very low-mass star and brown dwarf sample \\ Hidden water in carbon-dominated protoplanetary disks}

\author[0000-0001-8407-4020]{Aditya M. Arabhavi}
\affil{Kapteyn Astronomical Institute, Rijksuniversiteit Groningen, Postbus 800, 9700AV Groningen, The Netherlands}

\author[0000-0001-7455-5349]{Inga Kamp}
\affil{Kapteyn Astronomical Institute, Rijksuniversiteit Groningen, Postbus 800, 9700AV Groningen, The Netherlands}

\author[0000-0001-7591-1907]{Ewine F. van Dishoeck}
\affil{Leiden Observatory, Leiden University, PO Box 9513, 2300 RA Leiden, the Netherlands}
\affil{Max-Planck Institut f\"{u}r Extraterrestrische Physik (MPE), Giessenbachstr. 1, 85748, Garching, Germany}

\author[0000-0002-1493-300X]{Thomas Henning}
\affil{Max-Planck-Institut f\"{u}r Astronomie (MPIA), K\"{o}nigstuhl 17, 69117 Heidelberg, Germany}

\author[0000-0002-6592-690X]{Hyerin Jang}
\affil{Department of Astrophysics/IMAPP, Radboud University, PO Box 9010, 6500 GL Nijmegen, The Netherlands}

\author[0000-0002-0101-8814]{Valentin Christiaens}
\affil{Institute of Astronomy, KU Leuven, Celestijnenlaan 200D, 3001 Leuven, Belgium}
\affil{STAR Institute, Universit\'e de Li\`ege, All\'ee du Six Ao\^ut 19c, 4000 Li\`ege, Belgium}

\author[0000-0002-1257-7742]{Danny Gasman}
\affil{Institute of Astronomy, KU Leuven, Celestijnenlaan 200D, 3001 Leuven, Belgium}

\author[0000-0001-7962-1683]{Ilaria Pascucci}
\affil{Department of Planetary Sciences, University of Arizona; 1629 East University Boulevard, Tucson, AZ 85721, USA}

\author[0000-0002-8545-6175]{Giulia Perotti}
\affil{Max-Planck-Institut f\"{u}r Astronomie (MPIA), K\"{o}nigstuhl 17, 69117 Heidelberg, Germany}

\author[0000-0002-4022-4899]{Sierra L. Grant}
\affil{Earth and Planets Laboratory, Carnegie Institution for Science, 5241 Broad Branch Road, NW, Washington, DC 20015, USA}

\author[0000-0002-5971-9242]{David Barrado}
\affil{Centro de Astrobiolog\'ia (CAB), CSIC-INTA, ESAC Campus, Camino Bajo del Castillo s/n, 28692 Villanueva de la Ca\~nada, Madrid, Spain}

\author[0000-0001-9818-0588]{Manuel G\"udel}
\affil{Dept. of Astrophysics, University of Vienna, T\"urkenschanzstr. 17, A-1180 Vienna, Austria}
\affil{ETH Z\"urich, Institute for Particle Physics and Astrophysics, Wolfgang-Pauli-Str. 27, 8093 Z\"urich, Switzerland}

\author{Pierre-Olivier Lagage}
\affil{Universit\'e Paris-Saclay, Universit\'e Paris Cit\'e, CEA, CNRS, AIM, F-91191 Gif-sur-Yvette, France}

\author[0000-0001-8876-6614]{Alessio Caratti o Garatti}
\affil{INAF – Osservatorio Astronomico di Capodimonte, Salita Moiariello 16, 80131 Napoli, Italy}
\affil{Dublin Institute for Advanced Studies, 31 Fitzwilliam Place, D02 XF86 Dublin, Ireland}

\author{Fred Lahuis}
\affil{SRON Netherlands Institute for Space Research, PO Box 800, 9700 AV, Groningen, The Netherlands}

\author[0000-0002-5462-9387]{L. B. F. M. Waters}
\affil{Department of Astrophysics/IMAPP, Radboud University, PO Box 9010, 6500 GL Nijmegen, The Netherlands}
\affil{SRON Netherlands Institute for Space Research, Niels Bohrweg 4, NL-2333 CA Leiden, the Netherlands}

\author[0000-0001-8240-978X]{Till Kaeufer}
\affil{Space Research Institute, Austrian Academy of Sciences, Schmiedlstr. 6, A-8042, Graz, Austria}
\affil{Kapteyn Astronomical Institute, Rijksuniversiteit Groningen, Postbus 800, 9700AV Groningen, The Netherlands}
\affil{SRON Netherlands Institute for Space Research, Niels Bohrweg 4, NL-2333 CA Leiden, the Netherlands}
\affil{Institute for Theoretical Physics and Computational Physics, Graz University of Technology, Petersgasse 16, 8010 Graz, Austria}
\affil{Department of Physics and Astronomy, University of Exeter, Exeter EX4 4QL, UK}

\author[0000-0003-0386-2178]{Jayatee Kanwar}
\affil{Kapteyn Astronomical Institute, Rijksuniversiteit Groningen, Postbus 800, 9700AV Groningen, The Netherlands}
\affil{Space Research Institute, Austrian Academy of Sciences, Schmiedlstr. 6, A-8042, Graz, Austria}
\affil{TU Graz, Fakultät für Mathematik, Physik und Geodäsie, Petersgasse 16 8010 Graz, Austria}

\author[0000-0001-9526-9499]{Maria Morales-Calder\'on}
\affil{Centro de Astrobiolog\'ia (CAB), CSIC-INTA, ESAC Campus, Camino Bajo del Castillo s/n, 28692 Villanueva de la Ca\~nada,
Madrid, Spain}
 
\author[0000-0002-6429-9457]{Kamber Schwarz}
\affil{Max-Planck-Institut f\"{u}r Astronomie (MPIA), K\"{o}nigstuhl 17, 69117 Heidelberg, Germany}
 
\author[0000-0003-0330-1506]{Andrew D. Sellek}
\affil{Leiden Observatory, Leiden University, PO Box 9513, 2300 RA Leiden, the Netherlands}

\author[0000-0002-1103-3225]{Beno\^{i}t Tabone}
\affil{Universit\'e Paris-Saclay, CNRS, Institut d’Astrophysique Spatiale, 91405, Orsay, France}

\author[0000-0002-7935-7445]{Milou Temmink}
\affil{Leiden Observatory, Leiden University, PO Box 9513, 2300 RA Leiden, the Netherlands}

\author[0000-0002-3135-2477]{Marissa Vlasblom}
\affil{Leiden Observatory, Leiden University, PO Box 9513, 2300 RA Leiden, the Netherlands}



\begin{abstract}

Infrared observations of the inner disks around very low-mass stars (VLMS, $<$0.3$\,M_{\odot}$) have revealed a carbon-rich gas composition in the terrestrial planet-forming regions. Contrary to the typically water-rich T Tauri disk spectra, only two disks around VLMS have been observed to be water-rich among more than ten VLMS disks observed so far with JWST/MIRI. In this letter, we systematically search for the presence of water and other oxygen-bearing molecules in the JWST/MIRI spectra of ten VLMS disks from the MIRI mid-INfrared Disk Survey (MINDS). In addition to the two previously reported detections of water emission in this VLMS sample, we detect water emission in the spectra of three other sources and tentatively in one source, and we provide strong evidence for water emission in the remaining disks in the MINDS sample, most of which have bright emission from carbon-bearing molecules. We show that the \ch{C_2H_2} emission is much stronger than that of water for sources with low luminosities, and the hydrocarbons outshine the water emission in such conditions. \hl{We propose that the appearance of} water-rich vs. hydrocarbon-rich spectra is related to the location of the water reservoir in the disk relative to the main hydrocarbon reservoir. Our findings indicate that the terrestrial planet forming regions in VLMS disks \hl{have high carbon-to-oxygen ratios} (C/O$>$1), but can still harbor ample water similar to those in the T Tauri disks.

\end{abstract}



\section{Introduction}

Very low-mass stars (together with brown dwarfs, referred to as VLMS, $M_*$$<$ 0.3$\,M_{\odot}$) host most of the terrestrial planets detected so far (\citealt{2023AJ....165..265M}, \citealt{2021A&A...653A.114S}). Such planets inherit the planet-forming ingredients from the gas and dusk disks around VLMS in which they form. Rapid inward transport of pebbles \citep[e.g.][]{2013A&A...554A..95P,2024A&A...689A.236S} and carbonaceous grain and PAH destruction (e.g. \citealt{2010AdSpR..46...44K}, \citealt{2017ApJ...845...13A}, \citealt{2023NatAs...7..805T}) strongly influence the oxygen and carbon elemental abundances in the planet-forming regions. Models suggest that the evolution of temperate rocky planets, such as TRAPPIST-1, is sensitive to the carbon and oxygen budget of the primary atmosphere inherited from such disks (e.g. \citealt{2018AJ....155..195W}, \citealt{2024NatCo..15.8374K}). Consequently, it is important to characterize the chemical composition of the planet-forming regions around VLMS.

Observations at infrared wavelengths can be used to probe the chemical composition of the inner regions of planet-forming disks. \textit{James Webb Space Telescope} (JWST) observations have revealed rich hydrocarbon emission from disks around VLMS, indicating high carbon-to-oxygen ratios (C/O) compared to the T Tauri disks (\citealt{2023NatAs...7..805T}, \citealt{2024Sci...384.1086A}, \citealt{2024A&A...689A.231K}) as hinted at in lower resolution \textit{Spitzer}/IRS spectra (\citealt{2009ApJ...696..143P}, \citealt{2013ApJ...779..178P}). Arabhavi et al. (subm.) (further referred to as Paper 1) present the mid-infrared spectroscopy of the VLMS sample of the MIRI midINfrared Disk Survey (MINDS) GTO program (\citealt{2023FaDi..245..112K}, \citealt{2024PASP..136e4302H}), reporting that more than 85\% of the detected species contain carbon, and more than 50\% of the detected species are pure hydrocarbons. Some VLMS sources in this sample show very high hydrocarbon column densities that produce molecular pseudo-continuum emission (e.g., ISO-ChaI\,147: \citealt{2024Sci...384.1086A} and J16053215-1933159: \citealt{2023NatAs...7..805T}). Thermo-chemical models show that the observed hydrocarbons can form efficiently in the gas phase in the inner disks \citep[e.g.][]{2007ApJ...655L..49W,2024A&A...681A..22K} and that C/O ratios larger than unity increase the hydrocarbon abundances and column densities by several orders of magnitude (\citealt{2019ApJ...870..129W}, \citealt{2024A&A...689A.231K}). However, the models (e.g. \citealt{2024A&A...689A.231K}) also indicate that some water can be present in the inner disk gas even at high C/O. In addition, \citet{2023NatAs...7..805T}, \citet{2024Sci...384.1086A}, \citet{2024arXiv240806077K}, and \citet{2024arXiv241205535L} provide upper limits on water column densities in the carbon-rich sources that indicate that considerable amounts of water could still be hidden in the forest of hydrocarbon emission.

While only tentative detections of water were made with \textit{Spitzer} in two VLMS disks \citep{2013ApJ...779..178P}, \citet{2023ApJ...959L..25X} presented the JWST MIRI spectrum of a VLMS disk, Sz114 (0.17\,$M_{\odot}$), which showed a forest of water lines including the hot ro-vibrational band at $\sim$6.6\,$\mu$m and the cold pure rotational lines up to 28\,$\mu$m. \ch{C_2H_2} was the only hydrocarbon detected in that spectrum. Perotti et al. (subm.) and Morales-Calder\'on et al. (in prep.) report detections of pure rotational and ro-vibrational water lines, respectively, in the MIRI spectra of the brown dwarf disks, J04381486+2611399 (see also \citealt{2013ApJ...779..178P}) and Cha\,H$\rm \alpha$\,1. The latter is so far the only \hl{VLMS disk} where several hydrocarbons have been detected along with water. 

In this letter, we systematically search for oxygen-bearing molecules, more importantly water, in the MIRI spectra of VLMS disks in the MINDS program, and comment on the physical location of the water and hydrocarbon emitting reservoirs. We show that the hydrocarbon-rich inner disk regions of VLMS disks might not be as `dry' as suggested in previous analyses, and instead can contain water with similar \hl{continuum normalized water emission strengths} as in T Tauri disks which has important implications for planet formation and evolution.

\section{Observations and models}

The MINDS VLMS sample consists of ten sources with stellar masses ranging from 0.02\,$\rm M_{\odot}$ to 0.14\,$\rm M_{\odot}$ spanning three star-forming regions and one moving group (see Table\,\ref{tab:sources}, PID: 1282 and 1270). All \hl{of} our sources were obtained with target acquisition, and observed with all three MIRI Medium Resolution Spectroscopy \hl{(MRS) }gratings using a four-point dither pattern. We used the MINDS hybrid data reduction pipeline \citep{2024ascl.soft03007C} \hl{to obtain the reduced spectra. Further, we used the procedure described in \citet{Temmink2024} to determine the continuum for each source. The molecular continua are included in the molecular flux measurements.} More details about the observations, data reduction, \hl{continuum determination,} and the source properties of the sample can be found in Paper 1. Here, we limit our investigation to the detections of oxygen-bearing species and the evidence for the presence of underlying water emission in case of non-detections. 

We use ProDiMo 0D slab models presented in \citet{2024Sci...384.1086A} and HITRAN molecular spectroscopic data \citep{2022JQSRT.27707949G} throughout the letter. These models assume a slab of gas in local thermodynamic equilibrium (LTE) and characterized by column density, temperature, and emitting area. Mutual line opacity overlap of lines of the same species is also included. Our analysis of water detections indicates column densities similar to those found in the T Tauri disk spectra \hl{could be present in our sample, as will be explained later in Sect.\,\ref{sec:slabpeak}}, but the precise retrieval of the temperatures and column densities is beyond the scope of this letter. All spectra and the quantities derived from them are normalized to a distance of 150\,pc. 

\section{Detections of CO, CO$_2$, $^{13}$CO$_2$, and OH}

\hl{Detections of \ch{CO_2}, and \ch{^{13}CO_2} are adopted from Paper\,I and are included here for context (summarized in Table\,\ref{tab:sources}).} One of the main O-bearing species, \ch{CO_2}, is readily detected in VLMS because it has a prominent $Q$-branch (see Paper 1 and also models in Section 5.3). \ch{CO_2} and \ch{^{13}CO_2} are firmly detected in 9 and 8 of the 10 sources in our sample, respectively. The \ch{^{13}CO_2} detection indicates an optically thick reservoir of \ch{CO_2} ($\gtrsim$10$^{18}$\,cm$^{-2}$, e.g. see \citealt{2025A&A...693A.278V}, \citealt{Grant2023}).

\ch{CO} traditionally locks up a significant amount of oxygen because of its strong bond, but is not easy to detect in our sources. It emits shortwards of 5.5\,$\mu$m with a few high-$J$ $P-$branch lines of the $\nu$=1-0 and $\nu$=2-1 transitions. Although there is infrared excess from the disk at these wavelengths in all of our sources (Paper 1), the spectral features are dominated by the stellar photospheric absorption lines of \ch{CO} and \ch{H_2O} in seven of the ten sources in our sample (Figs.\,\ref{fig:hiddenwater_abs} and \ref{fig:coabs}, also see \citealt{2024Sci...384.1086A}, \citealt{2024A&A...689A.231K}). Quantifying the \ch{CO} disk emission signal requires proper modeling of the photospheres of the host star or brown dwarf which is outside the scope of this study. In the remaining three sources, we detect CO emission in two sources only, J1605 \citep[][Kanwar et al. subm.]{2023NatAs...7..805T} and J1558 (see Fig.\,\ref{fig:coemission}). \hl{The other source, NC1, shows infrared continuum and line excess shortwards of 5.5\,$\mu$m, but does not show \ch{CO} line emission. This will be discussed further in Morales-Calder\'on et al. (in prep.).}

\ch{OH} emission is detected in J1558 beyond 16\,$\mu$m and up to 25.3\,$\mu$m \hl{(see App.\,\ref{app:waterlongwave} for more details)} and tentatively in J0438 (Perotti et al. subm.).  

In summary, there are indications of O-bearing species in these disks, but apart from \ch{CO_2} they are hard to detect.

\begin{table}[h!]
    \centering
    \caption{Summary of the detections of emission of oxygen-bearing molecules from the inner disks of the MINDS VLMS sample. The colored boxes indicate the detections (green), non-detections (grey), and tentative detections (yellow) of oxygen-bearing species in the sample. $\rm H_2O_{rv}$ and $\rm H_2O_{ro}$ correspond to the ro-vibrational band and pure rotational line emission of water, respectively.}
    \begin{tabular}{P{12mm}P{8mm}P{8mm}P{8mm}P{6mm}P{0.5mm}P{0.5mm}P{0.5mm}P{0.5mm}P{0.5mm}P{0.5mm}}
    \hline\hline
         Short & SFR & SpTy & log$L_{*}$ & $M_{*}$ & \multirow{2}{*}{\begin{turn}{90}{$\rm H_2O_{rv}$}\end{turn}} & \multirow{2}{*}{\begin{turn}{90}{ $\rm H_2O_{ro}$}\end{turn}} & \multirow{2}{*}{\begin{turn}{90}{$\rm CO_2$}\end{turn}} & \multirow{2}{*}{\begin{turn}{90}{$\rm ^{13}CO_2$}\end{turn}} & \multirow{2}{*}{\begin{turn}{90}{$\rm OH$}\end{turn}} & \multirow{2}{*}{\begin{turn}{90}{ $\rm CO$}\end{turn}}\\
         name & & & [$L_{\odot}$] & [$M_{\odot}$] & & & & & & \\
        \hline
        J0438  & Tau & M7.25 & -2.70 & 0.05 & \myboxx{} & \mybox{} & \mybox{} & \mybox{} & \myboxxx{} & \myboxx{} \\
        J0439  & Tau & M6 & -1.00 & 0.12 & \myboxx{} & \mybox{} & \mybox{} & \mybox{} & \myboxx{} & \myboxx{} \\
         NC1  & ChaI  & M7.75 & -1.82 & 0.05 & \mybox{} & \myboxx{} & \mybox{} & \mybox{} & \myboxx{} & \myboxx{} \\
         NC9  & ChaI   & M5.5 & -1.42 & 0.08 & \myboxx{} & \myboxxx{} & \mybox{} & \mybox{} & \myboxx{} & \myboxx{} \\
         HKCha & ChaI   & M5.25 & -1.25 & 0.09 & \myboxx{} & \myboxx{} & \mybox{} & \mybox{} & \myboxx{} & \myboxx{} \\
         IC147 & ChaI   & M5.75 & -1.68 & 0.07  & \myboxx{} & \myboxx{} & \mybox{} & \mybox{} & \myboxx{} & \myboxx{} \\
         Sz28  & ChaI   & M5.25 & -1.37 & 0.08 & \myboxx{} & \myboxx{} & \mybox{} & \mybox{} & \myboxx{} & \myboxx{} \\
        TWA27 & TWA & M9 & -2.19 & 0.02 & \myboxx{} & \myboxx{} & \mybox{} & \mybox{} & \myboxx{} & \myboxx{} \\
         J1558 & UpSco & M4.5 & -1.35 & 0.14 & \myboxxx{} & \mybox{} & \myboxxx{} & \myboxx{} & \mybox{} & \mybox{} \\
         J1605 & UpSco  & M4.5 & -1.54& 0.13& \mybox{} & \myboxx{} & \mybox{} & \myboxxx{} & \myboxx{} & \mybox{} \\\hline
    \end{tabular}
    
    \justifying{\hspace{-0.5cm}\textbf{Notes.} The luminosities (not scaled to 150\,pc) and stellar masses are obtained from \citet{2023ASPC..534..539M}, updated based on the new GAIA EDR3 distances \citep{2021A&A...649A...1G}, except for NC1, TWA27, and J1605, for which the missing data are taken from \citet{2007ApJS..173..104L}, \citet{2024AJ....167..168M}, and \citet{2022A&A...663A..98T}, respectively. 2MASS source names are abbreviated above: J04381486+2611399, J04390163+2336029, J11071668-7735532, J11071860-7732516, J11074245-7733593, J11082650-7715550, J11085090-7625135, J12073346-3932539, J15582981-2310077, J16053215-1933159.}
    \label{tab:sources}
\end{table}

\begin{figure}[!ht]
    \centering
    \includegraphics[width=0.99\linewidth]{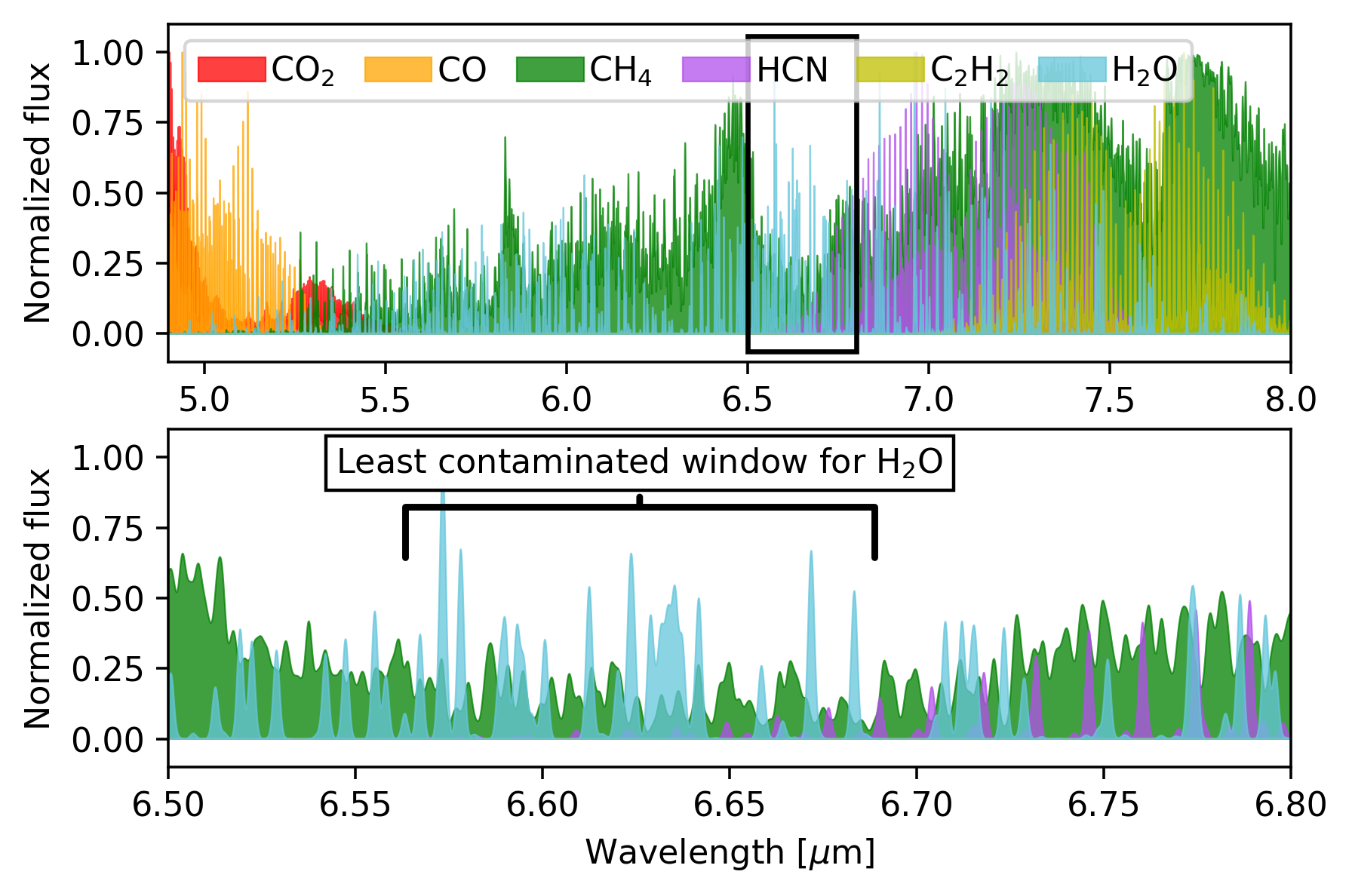}
    \caption{Simulated molecular emission at short wavelengths ($<$8\,$\mu$m). The top panel shows the peak-normalized slab model spectra of several molecules commonly detected in VLMS disks along with the ro-vibrational band of water \hl{at column densities and temperatures of 10$^{18.5}$\,cm$^{-2}$ and 525\,K, respectively. The column densities for \ch{CH_4} and \ch{C_2H_2} are 10$^{22}$\,cm$^{-2}$.} The black box shows the water dominant region (6.5-6.8\,$\mu$m), of which the bottom panel shows a zoom-in, highlighting the least contaminated region to identify the water ro-vibrational emission in observations. We assume a spectral resolving power of $R=3500$.}
    \label{fig:rovib}
\end{figure}

\section{Detections of water}

\begin{figure*}[!ht]
    \centering
    \includegraphics[width=\linewidth]{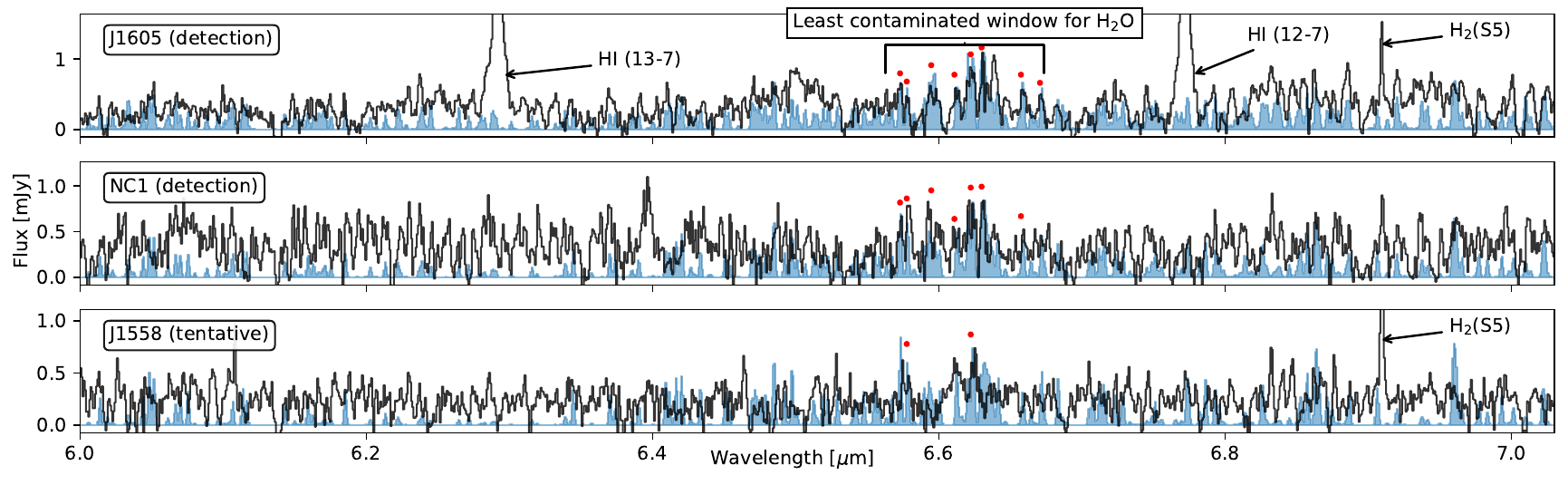}
    \caption{Detections of ro-vibrational bands of water emission in J1605 (top), NC1 (middle), and J1558 (bottom). The continuum subtracted spectra are shown in black and the LTE water slab models are shown in blue. The slab model parameters (temperature, column density, and emitting radius) are chosen to roughly match the observed spectra and are not obtained from fits: 525\,K, 10$^{20}$\,cm$^{-2}$, 0.02\,au for J1605; 525\,K, 10$^{19}$\,cm$^{-2}$, 0.02\,au for NC1; 525\,K, 10$^{18.33}$\,cm$^{-2}$, 0.03\,au for J1558. \hl{The red dots show the detected water lines in the least contaminated window.}}
    \label{fig:hiddenwater_rovib}
\end{figure*}

\begin{figure*}[!ht]
    \centering
    \includegraphics[width=\linewidth]{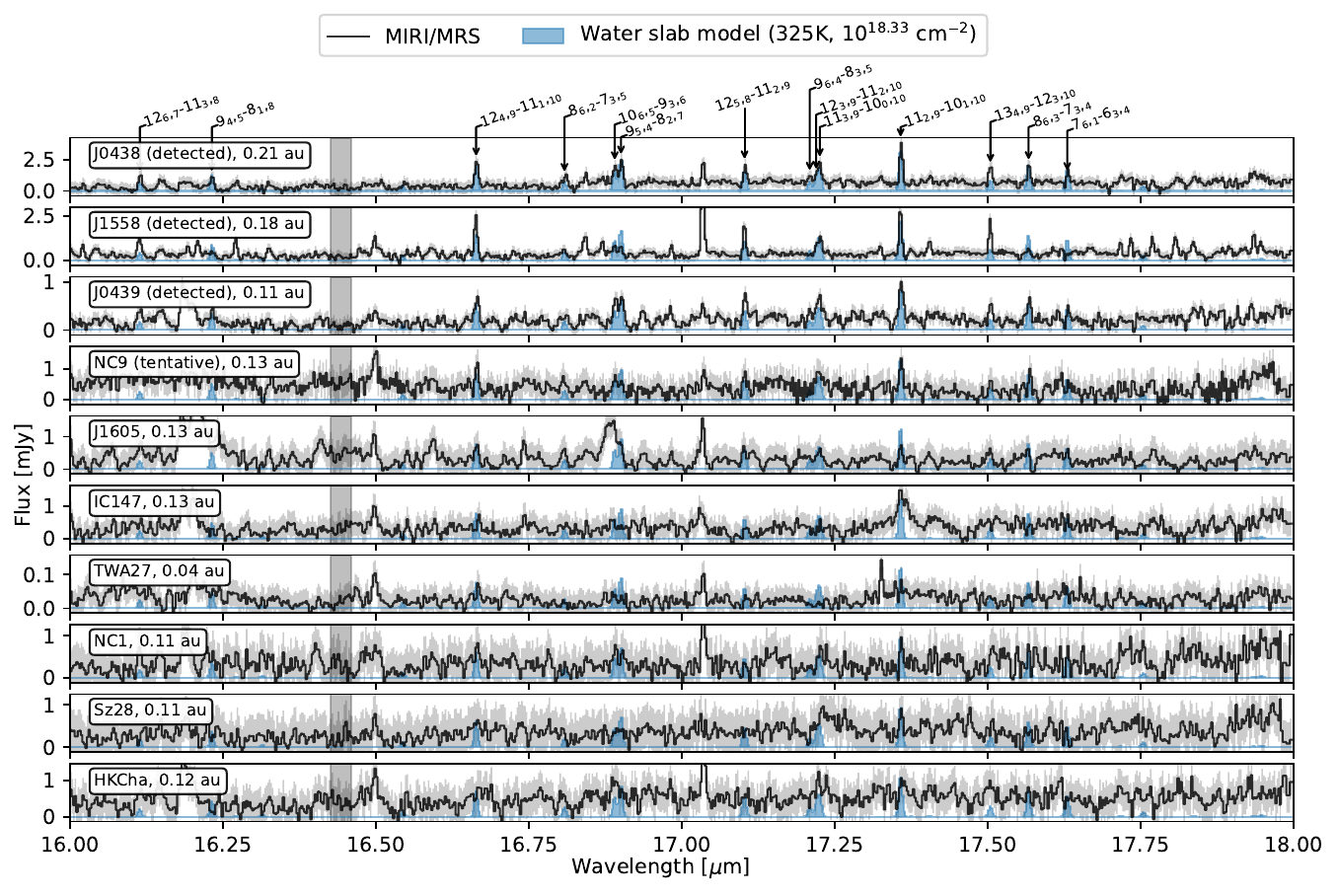}
    \caption{Detections and non-detections of pure rotational line emission of water. The continuum subtracted spectra are shown in black and the water slab models in blue. The slab model parameters are arbitrarily chosen for a visual match with the observed spectra. The temperature and column densities of the models are the same across all the panels, while the emitting area is scaled to match the flux level. The vertical gray area in each panel shows the wavelength region used to calculate the noise level. The light gray shaded region across the spectra shows the $\pm$3$\sigma$ noise level obtained in this way. The identified water line transitions are indicated at the top of the figure as $J_{K_a\,K_c}-J'_{K_{a}'\,K_{c}'}$, where $J$, $K_a$, $K_c$ are the total angular momentum quantum number, and the projections of the angular momentum on the principal axes.}
    \label{fig:hiddenwater}
\end{figure*}

Given the presence of \ch{CO_2} in most of our sources, the natural question is how much oxygen is contained in the water. \hl{The MIRI wavelength range covers the ro-vibrational band of water ($\nu_2$=1-0) at short wavelengths ($\sim$5-8\,$\mu$m) and the pure rotational lines at longer wavelengths ($\gtrsim$12\,$\mu$m). The ro-vibrational region is often affected by stellar photospheric absorption, whereas the longer wavelengths are not. Moreover, the ro-vibrational emission is expected to arise from an inner and smaller region of the disk compared to the more extended regions of the pure rotational lines (e.g. see Fig. 3 of \citealt{2024A&A...689A.330T}). For clarity, we analyze these two regions separately in the following two subsections.}

\subsection{Ro-vibrational band emission}
\label{sec:rovib}

Three sources show clear excess line emission from the disk in these wavelengths. Since the VLMS spectra are extremely line-rich, we need to clearly define the detection criteria for the ro-vibrational band of water. The top panel of Fig.\,\ref{fig:rovib} shows slab model spectra at the short wavelengths of a number of molecules that are commonly detected at longer wavelengths in the VLMS spectra \hl{at conditions typical for VLMS disks}. We identify a very narrow wavelength range between 5 and 8\,$\mu$m, where water lines are stronger than lines of other molecules, and no $Q-$branches or peak emission of carbon-bearing molecules occur. The bottom panel of Fig.\,\ref{fig:rovib} zooms in on this wavelength range, and our \hl{visual} detection criterion is identifying at least three of the peaks highlighted there. The relative strengths of these peaks are very sensitive to the column density and non-LTE effects.

Among the three sources in which the disk emission dominates at the shorter wavelengths, NC1 has been reported to show ro-vibrational water emission (Morales-Calder\'on et al. in prep.). Additionally, we detect ro-vibrational water emission in the carbon-rich disk - J1605, and tentatively in J1558. Fig.\,\ref{fig:hiddenwater_rovib} summarizes these detections. To confirm this, we also compare these features to slab models of other species shown in Fig.\,\ref{fig:rovib} and did not find a match. \hl{A more quantitative evidence for these detections is provided in Sect.\,\ref{sec:crosscorr} using cross-correlations between observations and synthetic water spectra}. These three sources have bright hydrocarbon emission (Paper 1), and J1605 and J1558 also show \ch{CO} emission. We show water slab models with column densities and emitting area that visually match the observations, assuming a temperature of 525\,K (the temperature of \ch{C_2H_2} in J1605 reported by \citealt{2023NatAs...7..805T}\hll{, with a reasonable change in temperature water lines would still appear at the same wavelengths}). The observed molecular emission at these short wavelengths is a sum of disk emission and stellar absorption. Hence, a proper fit would not only require subtraction of emission of other molecules, but also modeling the stellar spectra in detail. The latter would be best done by including the near-IR spectroscopy as demonstrated by Patapis et al. (in prep.) for TWA27.

\subsection{Pure rotational line emission}
\label{sec:purerot}

Figure\,\ref{fig:hiddenwater} shows the continuum subtracted spectra of all sources in the 16-18\,$\mu$m wavelength region along with water slab model spectra with temperatures of 325\,K, column densities of $10^{18.5}$\,cm$^{-2}$, and arbitrary emitting areas to match the observed flux levels. These slab models are used only to identify the location of water emission, and are not based on fits. At longer wavelengths ($>$18\,$\mu$m), due to the very faint nature of these VLMS, the signal-to-noise ratio (SNR) drops significantly, making it challenging to detect pure rotational water lines in most of our sample. At wavelengths below 16\,$\mu$m, the emission of several organic molecules spectrally overlaps, and it is hard to identify possible water features (see Fig.\,\ref{fig:hiddenwater2} \hl{and the corresponding text in App.\,\ref{app:waterlongwave}}). Therefore, the detections we report are based on the 16-18\,$\mu$m wavelength range. \hl{Similar to that used by \citet{2023ApJ...959L..25X} for the case of Sz114.} We estimate the noise level as the standard deviation in the 16.426-16.46\,$\mu$m wavelength range. The measured noise is very likely overestimated due to the numerous weak lines of hydrocarbons spread across the spectrum in the hydrocarbon-rich sources (see Sect.\,\ref{sec:hidden}). We claim a water detection if we identify more than five water lines between 16-18\,$\mu$m above 3$\sigma$. \hl{A more quantitative evidence for detections is shown in Sect.\,\ref{sec:crosscorr} using cross-correlations.}

We clearly detect individual water lines in three sources \hl{(see Fig.\,\ref{fig:hiddenwater})}: J0438 (also Perotti et al. subm.), J1558, and J0439 (top three panels) and tentatively in NC9 (fourth panel). Although the ro-vibrational band is detected in J1605 and NC1, we do not detect pure rotational water lines in these sources \hl{(see Sects.\,\ref{sec:crosscorr}\,\&\,\ref{sec:emittingregions_tempgrad})}. \hl{For the more noisy spectra of NC9, J1605, IC147, TWA27, NC1, Sz28, and HKCha, changing the detection criteria of number of identified lines or the parameters of the slab models does not change the results.}

In summary, we have clear detections of \ch{H_2O} emission in half of our sources, either in ro-vibration or in pure rotational emission.

\section{Hidden water in the non-detections}
\label{sec:hidden}

The noise measured from the narrow spectral window leads to a signal-to-noise ratio (SNR) between 5-30 in our sources. However, the observational setup is designed for much better SNRs, the SNR reported in previous analyses and the exposure time calculator (ETC) are close to or above 100 (e.g. \citealt{2024Sci...384.1086A}, \citealt{2024A&A...689A.231K}). This clearly shows that the \hl{features in the spectral window} used to measure noise could largely be a forest of weak molecular emission. For instance, although J0438, J0439, and J1558 show clear water emission in the 16-18\,$\mu$m range (Fig.\,\ref{fig:hiddenwater}), the \ch{C_2H_2} and \ch{CO_2} $P-$ and $Q-$branch emission in J0439 and J1558 are stronger relative to J0438 and outshine the water emission lines at wavelengths shortward of 16\,$\mu$m, as shown in Fig.\,\ref{fig:hiddenwater2}. However, water remains detectable in J0438 even at $\sim$14\,$\mu$m. Moreover, the six sources with non-detections of pure rotational water emission are rich in hydrocarbon emission (Paper 1).

We hypothesize that the hydrocarbon emission in these six sources outshines the water emission. In the following subsections, we use two techniques to provide evidence for possible water emission in the VLMS spectra with non-detections in water: stacking of spectra with water non-detections and cross-correlations with water slab models.

\begin{figure*}[!ht]
    \centering
    \includegraphics[width=\linewidth]{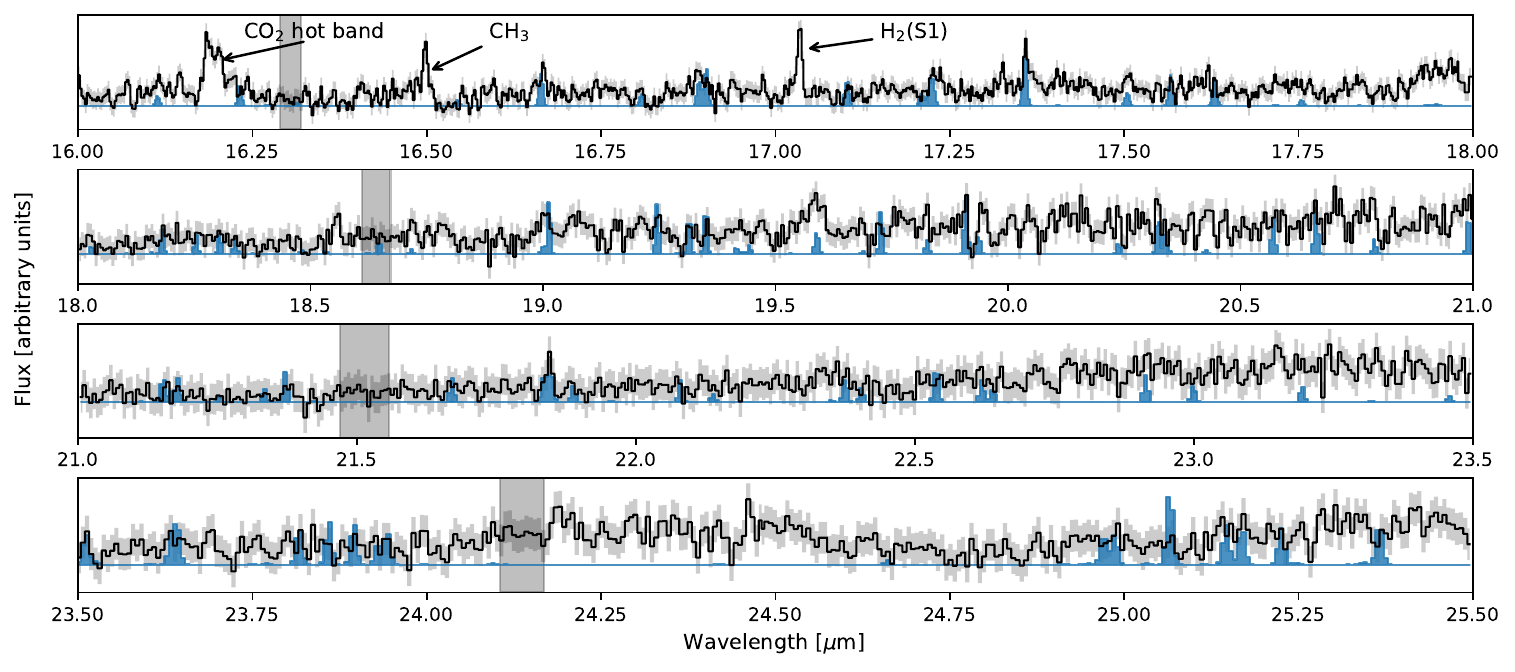}
    \caption{Stacked spectra of the sources in which pure rotational lines of water are not detected (black); the four panels show the 16 to 25.5\,$\mu$m region. The water slab model spectrum is shown in blue, and the gray shaded region around the spectrum shows the $\pm$3$\sigma$ noise level measured at wavelengths indicated by vertical gray boxes.}
    \label{fig:stacked}
\end{figure*}

\subsection{Stacking the spectra of non-detections}

Figure\,\ref{fig:stacked} shows the stacked spectrum of all sources in which we do not detect water lines (excluding NC9 with a tentative detection) for wavelengths beyond 16\,$\mu$m. For this analysis, we divide the wavelength range into four regions shown in the four panels of the figure. The spectrum ($F'$) in each panel is given by:
\begin{equation}
    F'(\lambda)=\frac{1}{n}\sum_{i=1}^{n}\frac{F_i(\lambda)}{\tilde{F_i}\sigma_i}
\end{equation}
here, $F$ is the continuum-subtracted spectrum shown in Figs.\,\ref{fig:hiddenwater},\,\ref{fig:hiddenwater20}-\ref{fig:hiddenwater26}, $\lambda$ is the wavelength, $\tilde{F}$ is the median of the continuum-subtracted spectrum in each panel, $\sigma$ is the variance measured in the gray shaded region shown in Fig.\,\ref{fig:stacked} in each panel, the subscript $i$ corresponds to individual sources, and $n$ is the total number of sources stacked, six in this case. Although stacking improves the signal-to-noise ratio, we still find residual \hl{features} in the stacked spectra which strongly suggest underlying real molecular signatures (we cannot neglect possible remaining data reduction artifacts such as fringes). In line with this, \ch{CH_3} and \ch{CO_2} features (which are detected in all the six sources) clearly show-up in the stacked spectrum. 

On the other hand, several features coincide with emission features in the water slab model. Using the same detection criteria as in \hl{Sect.\,\ref{sec:purerot}}, we clearly identify several water lines, with the most prominent ones at 17.36\,$\mu$m ($11_{2\,9}$-$10_{1\,10}$) and 21.85\,$\mu$m (blend of three lines: $9_{4\,6}$-$8_{1\,7}$, $11_{8\,3}$-$10_{7\,4}$, $10_{9\,2}$-$9_{8\,1}$). This again strongly suggests that there is water emission in some of these sources that do not have water line detections individually.

\subsection{Cross-correlation with slab models}
\label{sec:crosscorr}

\begin{figure*}[!ht]
    \centering
    \includegraphics[width=\linewidth]{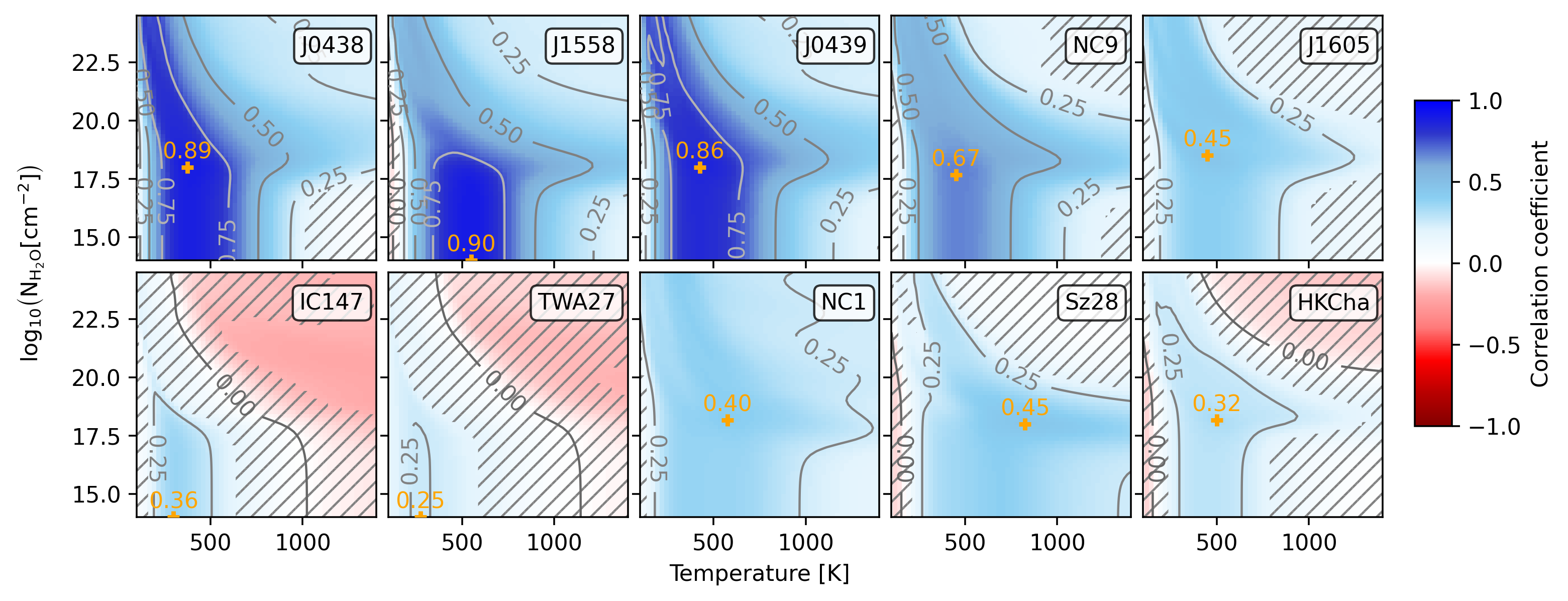}
    \caption{Cross-correlations of observed VLMS spectra with water slab models between 16.65 and 17.65\,$\mu$m for different column densities ($N$) and temperatures ($T$). The correlation coefficient varies from -1 (anti-correlation, in red) to +1 (correlation, in blue). The contours of correlation coefficients at 0.25, 0.5, and 0.75 are shown with dark gray lines. The parameter space where the statistical significance is low (i.e., the p-value is larger than 0.03) is hatched. The orange cross indicates the best correlation coefficient for each source.}
    \label{fig:water_corr}
\end{figure*}

Using a grid of water slab models (see \citealt{2024Sci...384.1086A} for the details of the slab models), we calculate the Pearson correlation coefficients (\texttt{scipy.stats.pearsonr}, \citealt{2020SciPy-NMeth}) between the normalized continuum-subtracted VLMS spectra and the normalized water slab model spectra between 16.65 and 17.65\,$\mu$m. Figure\,\ref{fig:water_corr} shows these correlation coefficients for the entire VLMS sample across a wide range of water column densities and temperatures. J1558, J0438, and J0439 are the sources in which we detect pure rotational water lines in Sect.\,\ref{sec:purerot}, and we indeed find strong statistically significant correlations with the water slab models. In the remaining sources, we find regions in the parameter space that show statistically significant correlation coefficients $\geq$0.25. These correlations are not as strong as in the sources with firmly detected rotational water lines. Notably, we find the weakest correlations in the sources with the highest number of organic molecule detections, IC147 and TWA27.

To confirm that such a cross-correlation analysis produces reliable results and does not produce false positives due to noise or emission from other molecules, we performed the same exercise on the spectrum of a debris disk devoid of water features and on synthetic spectra of other carbon-bearing molecules observed in our sample (see App.\,\ref{app:corr}). We conclude that the statistically significant correlation coefficients $\geq$0.25 indicate underlying water emission and are not false positives (App.\,\ref{app:corr}). 

In terms of slab model parameters, the strongest correlations are below column densities of $N=\rm 10^{18.5}\ cm^{-2}$ and temperatures $T$\,$<$1000\,K. It should be noted that the correlations only indicate the likelihood and do not indicate the best-fitting slab models. 

\hl{We also calculated the correlation coefficients for the ro-vibrational band of water (see App.\,\ref{app:corr}). We find strong statistically significant correlations for J1605 and NC1 where the ro-vibrational band emission of water is firmly detected, and strong statistically significant anti-correlations in the sources where water absorption by the stellar photosphere is detected. For J1558, which we classify as a tentative detection of the ro-vibrational band based on our visual detection criteria, the positive correlations are not statistically significant (see App.\,\ref{app:corr} for more details).}

\begin{figure}[!ht]
    \centering
    \includegraphics[width=0.99\linewidth]{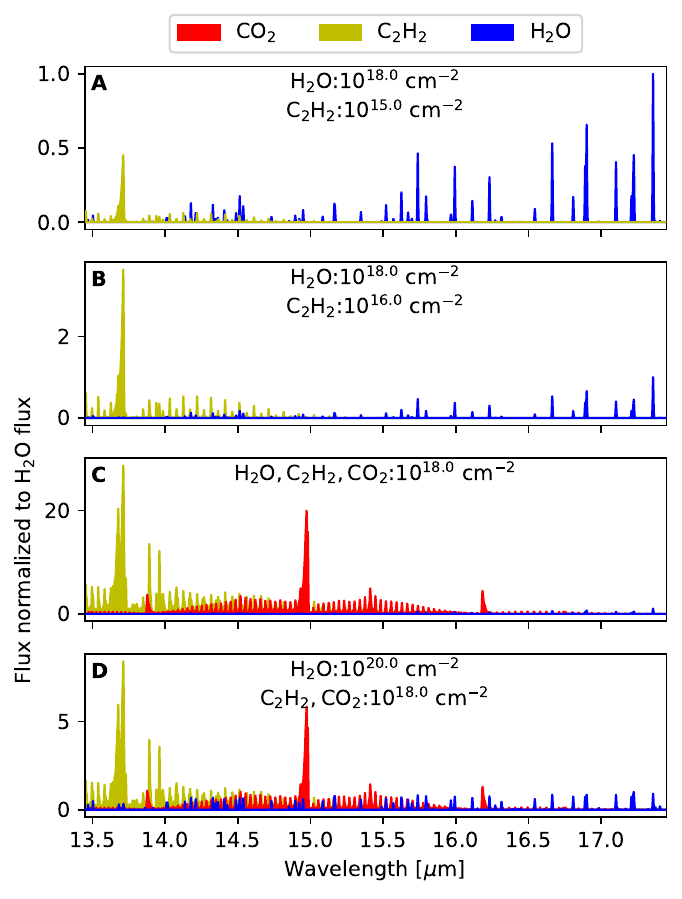}
    \caption{Comparison of \ch{C_2H_2}, \ch{CO_2}, and \ch{H_2O} flux levels for the different column densities (temperatures are fixed to 325\,K), normalized to the peak water flux in the shown wavelength range. Note the different vertical axis limits.}
    \label{fig:coolants}
\end{figure}

\subsection{Slab models and peak-to-continuum ratios}
\label{sec:slabpeak}

While we find evidence for weak water emission underneath the hydrocarbon-rich spectra in sources without firm water detection, \ch{CO_2} is prominently detected in all these sources. Therefore it is natural to investigate whether \hl{the clear \ch{CO_2} detections are} just due to the prominent $Q-$branch of \ch{CO_2}. Here, we show that water can still be present in significant abundances and could be easily missed because of its less prominent line emission spectrum.

Figure\,\ref{fig:coolants} compares slab model spectra of \ch{CO_2} and \ch{C_2H_2} at different emitting conditions with the slab model spectra of water to illustrate this. \hl{Panel A shows the slab spectra of \ch{H_2O} and \ch{C_2H_2} at 325\,K, and at low column densities of \ch{C_2H_2} (10$^{15}$\,cm$^{-2}$) relative to \ch{H_2O} (10$^{18}$\,cm$^{-2}$). While water is bright in Panel A, the \ch{C_2H_2} $Q-$branch flux increases almost linearly when the emission is optically thin, i.e. when the column densities are low ($\lesssim$10$^{17}$\,cm$^{-2}$), an increase in column density by an order of magnitude increases the flux by roughly an order of magnitude (see panel B). When these molecules are at the same column densities, the $P-$branch lines of \ch{C_2H_2} clearly outshine water  (panel C, also \ch{CO_2})}. It should be noted that the \ch{C_2H_2} column densities in the hydrocarbon-rich VLMS disks are, in fact, a few orders of magnitude larger than 10$^{18}$\,cm$^{-2}$, much larger than the upper limits for water column densities (\citealt{2024Sci...384.1086A}, \citealt{2023NatAs...7..805T}), and there are several other hydrocarbons detected in these VLMS disks (Paper I). This simple experiment reveals that in the presence of many abundant hydrocarbons the water could be hidden below the hydrocarbon emission. Panel D shows that even a water column density that is a factor of 100 larger would still only produce peak fluxes at the same level as the numerous $P-$branch lines. Moreover, the pure rotational water lines could also probe much lower temperatures than \ch{C_2H_2} (e.g., \citealt{2023ApJ...959L..25X}). This would then mean that even higher column densities and larger emitting areas of water could still be hidden in the forest of hydrocarbon emission (see App.\,\ref{app:fluxratios} for a more detailed look at the relative fluxes). \hl{This could explain the non-detections of pure rotational water lines in J1605 and NC1, where ro-vibrational water lines are clearly detected.}

\begin{figure}[!ht]
    \centering
    \includegraphics[width=0.99\linewidth]{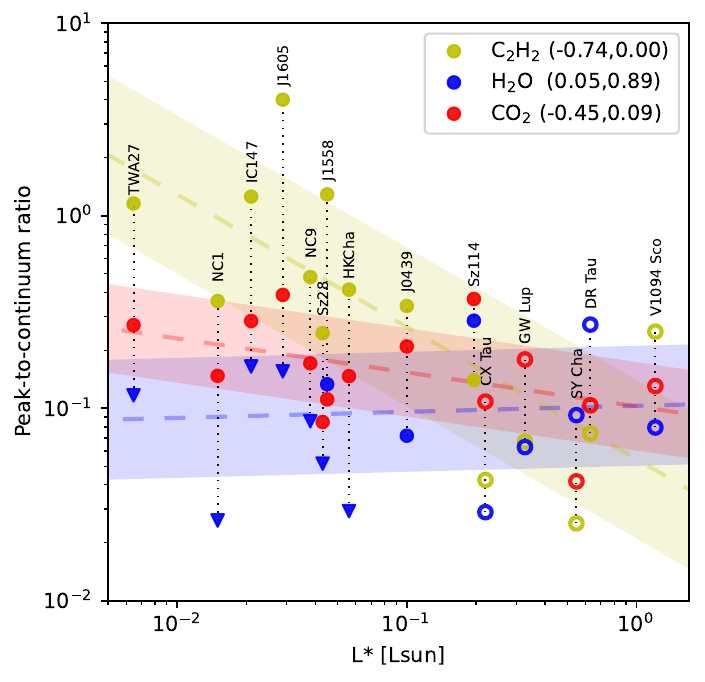}
    \caption{The peak-to-continuum ratio of \ch{C_2H_2}, \ch{CO_2}, and \ch{H_2O} in the MINDS sample (and Sz114 \citealt{2023ApJ...959L..25X}). The VLMS sources are indicated by filled circles, and the T Tauri sources by empty circles. The inverted triangles indicate upper limits (tentative or non-detections). The colors are same as in Fig.\,\ref{fig:coolants}. The dashed lines and the shaded regions show the linear fits to the peak-to-continuum ratios of the different species and the 1$\sigma$ standard deviation. The legend shows the correlation coefficients and the p-values. The water peak corresponds to the 17.36\,$\mu$m pure rotational line transition: 11$_{2,9}$-10$_{1,10}$, $E\rm_{up}$\,=\,$2432$\,K. References for the spectra of T Tauri sources: \citet{Grant2023}, \citet{Schwarz2024}, \citet{Temmink2024}, \citet{2025A&A...693A.278V}, Tabone et al. (in prep.).}
    \label{fig:line-cont}
\end{figure}

We can demonstrate that \ch{C_2H_2} emission outshines \ch{H_2O} by measuring the peak-to-continuum ratios of \ch{H_2O}, \ch{C_2H_2}, and \ch{CO_2} in the observed spectra across stellar luminosities (Fig.\,\ref{fig:line-cont}). The \ch{C_2H_2} peak-to-continuum ratio increases by close to an order of magnitude as the stellar luminosity decreases by two orders of magnitude, while the peak-to-continuum ratio of \ch{H_2O} does not show any statistically significant (p-value$<$0.03) change with stellar luminosity. Including or omitting the upper limits from sources with no firm water detection does not change our conclusions. The \ch{CO_2} peak-to-continuum ratio, with a marginally significant p-value of 0.09, lies between those of \ch{H_2O} and \ch{C_2H_2} indicating a tentative correlation. Hence, we conclude that \ch{C_2H_2} and \hl{possibly} \ch{CO_2} outshine water emission as the stellar luminosity decreases.

\hl{The column densities of \ch{H_2O} in the T Tauri disks estimated from the MIRI spectra span a wide range of 10$^{17}$-10$^{20}$\,cm$^{-2}$ (\citealt{2023A&A...679A.117G}, \citealt{2023ApJ...957L..22B}, \citealt{2023Natur.620..516P}, \citealt{2024ApJ...975...78R}, \citealt{Schwarz2024}, \citealt{2024A&A...689A.330T}, Temmink et al. subm.). The by-eye fits to the ro-vibrational and pure rotational water emission in our VLMS sample show that water column densities similar to the T Tauri disks could be hidden in these hydrocarbon-rich spectra. Detailed modeling of each source is required to accurately estimate the column densities.}

\section{Discussion}
\label{sec:discussion}

\subsection{Emitting regions and temperature gradients}
\label{sec:emittingregions_tempgrad}
\hl{If we adopt the high temperature slab models that visually match the ro-vibrational emission of water, cooler water slab components with larger emitting radii} are needed to match the pure rotational line fluxes in those sources (see App.\,\ref{app:waterlongwave} and Figs.\,\ref{fig:hiddenwater_rovib},\,\ref{fig:hiddenwater},\,\ref{fig:hiddenwater3}). These two temperature components could indicate a thermal gradient of water in VLMS disks, similar to Sz114 \citep{2023ApJ...959L..25X} and T Tauri disks (e.g., \citealt{2023A&A...679A.117G}, \citealt{2023ApJ...957L..22B}, \citealt{2024A&A...689A.330T}, \citealt{2024ApJ...975...78R}). Detailed multi-temperature modeling, considering the photosphere and other molecular emissions, is required to precisely quantify the water in these VLMS disks. 

\begin{figure}
    \centering
    \includegraphics[width=0.99\linewidth]{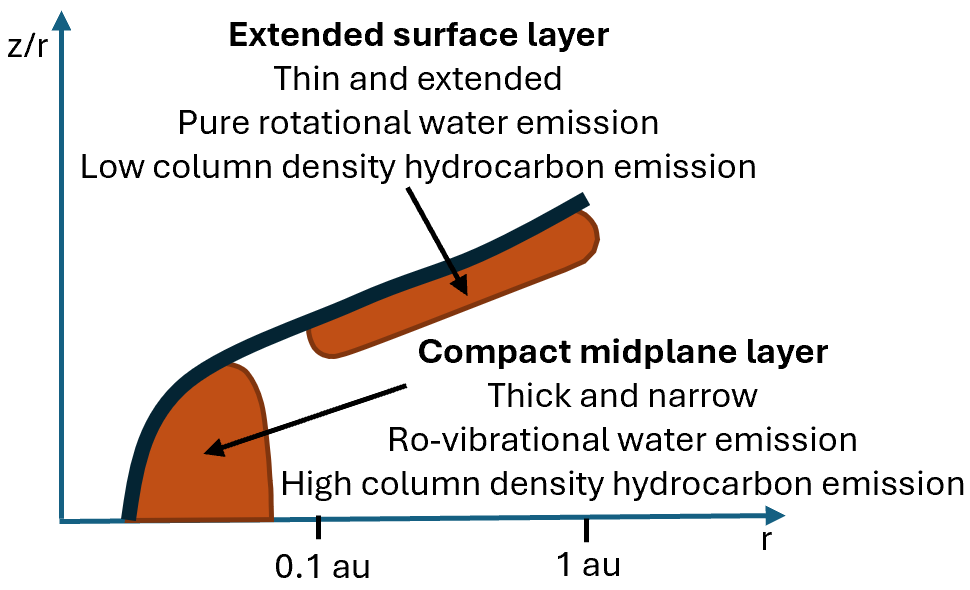}
    \caption{Sketch illustrating the different components of hydrocarbon and water emission.}
    \label{fig:diskstructure}
\end{figure}

\hl{Some disk models of VLMS }\citep[e.g.][]{2024A&A...689A.231K} propose two hydrocarbon reservoirs: a radially extended surface ($>$0.1\,au) and a compact midplane region ($<$0.1\,au), as sketched in Fig.\,\ref{fig:diskstructure}. Water is typically emitted from the disk surface \hl{with the ro-vibrational band emission arising radially closer to the central object than the more extended pure rotational emission} (\citealt{2018A&A...618A..57W}, \citealt{2022ApJ...930L..26B}).

\hl{The spectra can be water-rich in pure rotational lines when the \ch{C_2H_2} column densities are low ($\lesssim$10$^{17.5}$\,cm$^{-2}$, $>$375\,K). Sources like Sz114 and J0438 exhibit water-rich spectra, with water and \ch{C_2H_2} emission from extended layers ($>$0.1 au). In contrast, if the \ch{C_2H_2} column densities are large ($\gtrsim$10$^{17.5}$\,cm$^{-2}$), then the hydrocarbon emission can outshine the pure rotational water emission from the extended surface layers ($>$0.1 au). For example, sources such as IC147 and J1605 show hydrocarbon-rich spectra, with large column densities ($\gtrsim$10$^{20}$\,cm$^{-2}$) of hydrocarbons emitting from compact regions ($<$0.1\,au, \citealt{2024Sci...384.1086A}, \citealt{2023NatAs...7..805T}).  In such cases, the water rotational lines would be fainter than the $P-$branch lines of \ch{C_2H_2} (see App.\,\ref{app:fluxratios} for a more detailed look at the relative brightness of \ch{C_2H_2} and pure rotational lines of water).}

The presence of compact high column density ro-vibrational water emission in the hydrocarbon-rich disks (see Figs.\,\ref{fig:hiddenwater_rovib},\,\ref{fig:hiddenwater},\,\ref{fig:hiddenwater3}) could suggest co-spatial hot water and hydrocarbons in high C/O ($>$1) environments, aligning with predictions (\citealt{2024A&A...681A..22K}).

The emitting radii of pure rotational lines (Fig.\,\ref{fig:hiddenwater}) are similar to the typical water iceline in models of disks around such faint stars (e.g. \citealt{2017A&A...601A..44G}). The smaller emitting radii of abundant hydrocarbons ($<$0.1\,au) compared to the rotational water lines could indicate a strong radial gradient in the volatile C/O ratio.

\subsection{Implications for formation and evolution of planets}
\label{sec:implications}

The composition of the inner disk largely shapes the terrestrial planets that form in these regions. The diverse hydrocarbons detected in the inner regions of these disks indicate that the volatile C/O ratio is larger than unity (e.g., \citealt{2023NatAs...7..805T}, \citealt{2024Sci...384.1086A}, \citealt{2024A&A...689A.231K}), which is higher than the typical volatile C/O ratio expected in the inner regions of the T Tauri disks ($\sim$0.45). Although planets that form in the \hl{inner disks around VLMS} could encounter highly \hl{carbon-rich} environments, our findings show that the planet-forming material is not necessarily dry. The planets around VLMS could inherit water abundances similar to that around the Sun-like stars given the large column densities and emitting area of water vapor that we show could be present in these disks. Further, a radial gradient in C/O ratio could lead to different planet evolution in the inner ($<$0.1\,au) and outer disks.

Compared to the rest of our sources, TWA27 is expected to be the oldest source ($\sim$10\,Myr, \citealt{2006A&A...459..511B}, \citealt{2015MNRAS.454..593B}) and shows the weakest evidence for water (Fig.\,\ref{fig:water_corr}). Moreover, the emitting radius from the by-eye fits of the pure rotational lines of TWA27 is more than a factor of two smaller than the rest of the sample. As suggested by models such as \citet{2023A&A...677L...7M}, the rapid transport ($<$2-3\,Myr) in the midplane could have depleted oxygen in TWA27. Planets that form at late stages could evolve under water-poor conditions. \hl{Alternatively, if the high gaseous C/O ratio is due to the destruction of carbonaceous grains or polycyclic aromatic hydrocarbons (as suggested for J1605, \citealt{2023NatAs...7..805T}), then terrestrial planets that form from such material could be depleted in carbon.}

\section{Conclusions}
Previous analyses of hydrocarbon-rich VLMS disk spectra revealed abundant hydrocarbon emission with an absence of water emission. We provide evidence for water in these disks, showing that inner regions with high C/O ratios can still harbor substantial water column densities. 

We detect several oxygen-bearing molecules in the VLMS sample of the MINDS program, such as \ch{CO}, \ch{CO_2}, \ch{^{13}CO_2}, \ch{OH}, and \ch{H_2O}, including new detections of the ro-vibrational emission of water in J1605, and tentatively in J1558, and the pure rotational line emission of water in J0439 and J1558, and tentatively in NC9. \hl{We find tentative evidence for a thermal gradient in the water emission in the sources which show ro-vibrational water emission, but detailed modeling is required to draw robust conclusions.}

\hl{We show that \ch{C_2H_2} becomes brighter as the luminosity of the central object decreases. Our results demonstrate that bright \ch{C_2H_2}, along with possibly \ch{CO_2} and other hydrocarbons, can dominate the spectra and outshine water emission in VLMS disks.}

Our findings suggest a nuanced picture of planet formation, where \hl{the inner disk} regions could foster high C/O ratios while still retaining significant water content. Any gradients in volatile C/O ratios across the inner disk region further emphasize the potential for varied evolutionary pathways for planets forming within these systems.

As a next step, the water abundances have to be quantified. This would not only provide an estimate of the water budget for planet formation, but would also provide strong constraints on the efficiency of the inward transport processes in these disks. To quantify the abundance of water and characterize its temperature gradient, proper modeling of the stellar photospheres is required for sources that are dominated by absorption lines at shorter wavelengths $<$7\,$\mu$m to investigate the presence of ro-vibrational \ch{CO} and \ch{H_2O} emission from the disk. Moreover, non-LTE effects should be studied for the disk environments around VLMSs. Moreover, a comprehensive modeling of hydrocarbon emission is required, which in turn requires more molecular spectroscopic data such as \ch{^{13}CCH_2}.

\section*{Acknowledgments}
This work is based on observations made with the NASA/ESA/CSA James Webb Space Telescope. The data were obtained from the Mikulski Archive for Space Telescopes at the Space Telescope Science Institute, which is operated by the Association of Universities for Research in Astronomy, Inc., under NASA contract NAS 5-03127 for JWST. These observations are associated with program \#1282. The specific observations analyzed can be accessed via \dataset[doi: 10.17909/zrxj-0j97]{https://doi.org/10.17909/zrxj-0j97}. The following National and International Funding Agencies funded and supported the MIRI development: NASA; ESA; Belgian Science Policy Office (BELSPO); Centre Nationale d’Etudes Spatiales (CNES); Danish National Space Centre; Deutsches Zentrum fur Luft- und Raumfahrt (DLR); Enterprise Ireland; Ministerio De Econom\'ia y Competividad; Netherlands Research School for Astronomy (NOVA); Netherlands Organisation for Scientific Research (NWO); Science and Technology Facilities Council; Swiss Space Office; Swedish National Space Agency; and UK Space Agency. I.K., A.M.A., and E.v.D. acknowledge support from grant TOP-1 614.001.751 from the Dutch Research Council (NWO). E.v.D. acknowledges the support from the Danish National Research Foundation through the Center of Excellence ``InterCat'' (DNRF150). E.v.D., A.D.S, M.T., and M.V. acknowledge support from the ERC grant 101019751 MOLDISK. T.H. and K.S. acknowledge support from the European Research Council under the Horizon 2020 Framework Program via the ERC Advanced Grant Origins 83 24 28. V.C. acknowledges funding from the Belgian F.R.S.-FNRS. D.G. thanks the Belgian Federal Science Policy Office (BELSPO) for the provision of financial support in the framework of the PRODEX Programme of the European Space Agency (ESA). G.P. gratefully acknowledges support from the Max Planck Society. A.C.G. acknowledges support from PRIN-MUR 2022 20228JPA3A “The path to star and planet formation in the JWST era (PATH)” funded by NextGeneration EU and by INAF-GoG 2022 “NIR-dark Accretion Outbursts in Massive Young stellar objects (NAOMY)” and Large Grant INAF 2022 “YSOs Outflows, Disks and Accretion: towards a global framework for the evolution of planet forming systems (YODA)”. D.B. has been funded by Spanish MCIN/AEI/10.13039/501100011033 grants PID2019-107061GB-C61 and No. MDM-2017-0737. I.K., J.K., and T.K. acknowledge funding from H2020-MSCA-ITN-2019, grant no. 860470 (CHAMELEON). B.T. acknowledges support from the Programme National “Physique et Chimie du Milieu Interstellaire” (PCMI) of CNRS/INSU with INC/INP co-funded by CEA and CNES. I.P. acknowledges partial support by NASA under agreement No. 80NSSC21K0593 for the program ``Alien Earths.''

%

\vspace{5mm}
\facilities{JWST/MIRI}


\software{\href{https://prodimo.iwf.oeaw.ac.at/}{ProDiMo} \citep{2016A&A...586A.103W},
          \href{https://gitlab.astro.rug.nl/prodimo/prodimopy}{prodimopy} \citep{2024Sci...384.1086A},
          scipy \citep{2020SciPy-NMeth}, matplotlib \citep{Hunter:2007}, numpy \citep{harris2020array}, SpectRes \citep{carnall2017}
          }



\appendix

\section{The short wavelengths: Emission and stellar photospheric absorption lines of CO and H2O}
\begin{figure}[!ht]
    \centering
    \includegraphics[width=\linewidth]{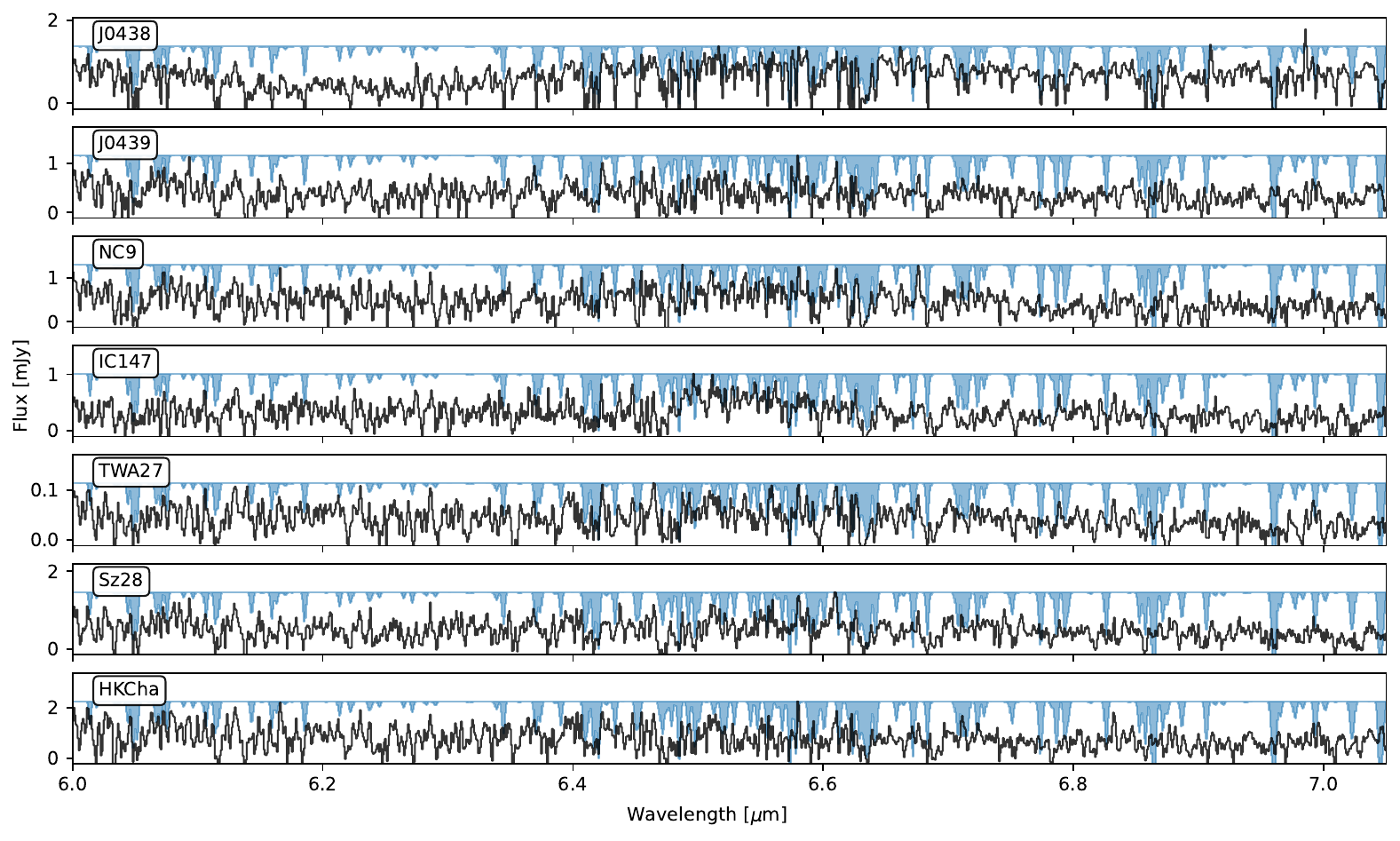}
    \caption{Overview of the stellar photospheric absorption features of water observed in the sample. The continuum subtracted spectra are shown in black and the water absorption models are shown in blue. The stellar photospheric absorption features in the spectra of IC147, Sz28, and J0438 have been reported in \citet{2024Sci...384.1086A}, \citet{2024A&A...689A.231K}, and Perotti et al. (subm.) respectively.}
    \label{fig:hiddenwater_abs}
\end{figure}

\begin{figure}[!ht]
    \centering
    \includegraphics[width=\linewidth]{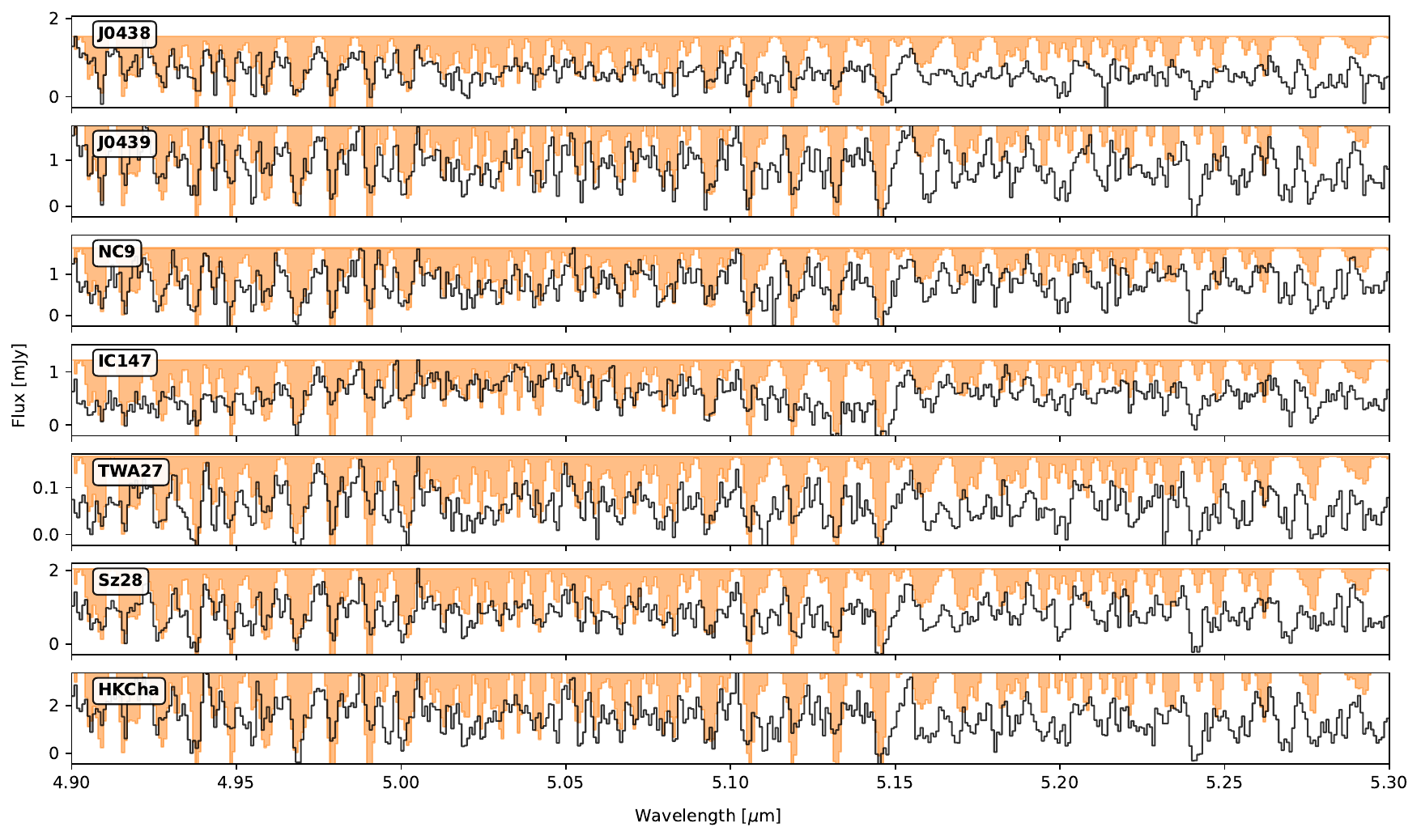}
    \caption{Same as Fig.\,\ref{fig:hiddenwater_abs}, but stellar photospheric absorption of \ch{CO} is shown in orange.}
    \label{fig:coabs}
\end{figure}

\begin{figure}[!ht]
    \centering
    \includegraphics[width=\linewidth]{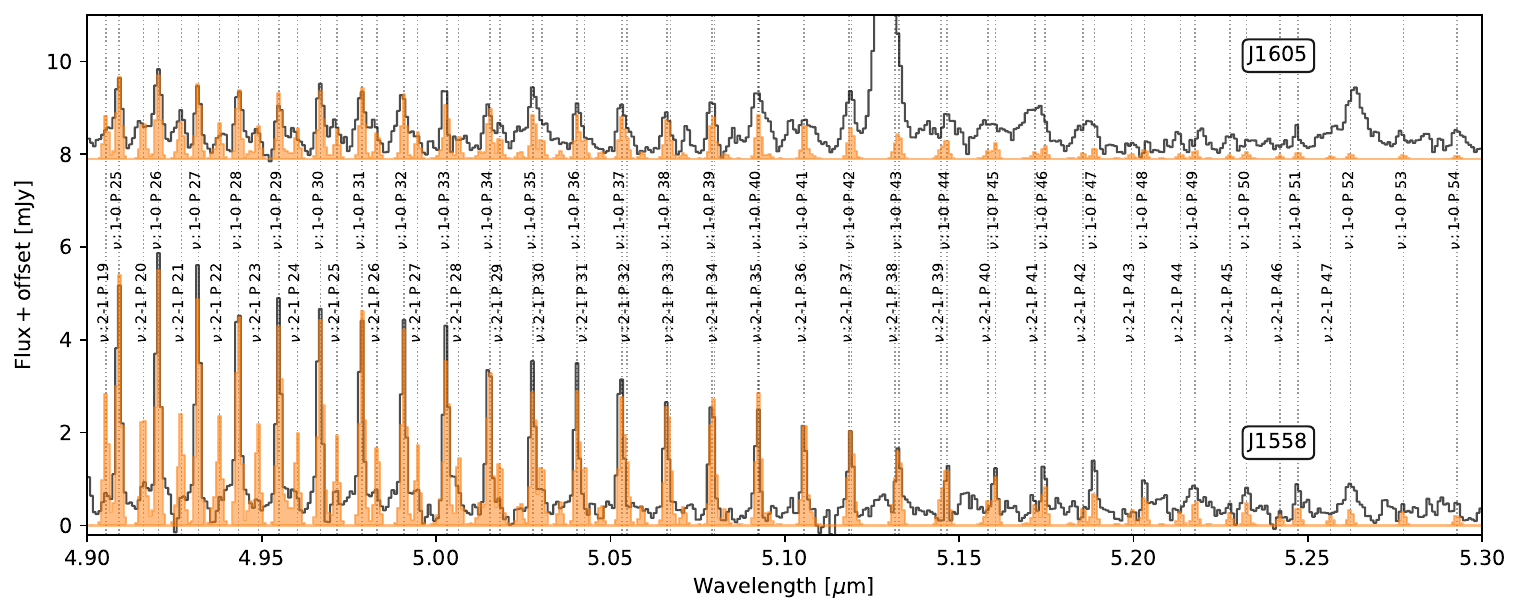}
    \caption{Detections of \ch{CO} emission in J1605 and J1558. The continuum subtracted spectra are shown in black and the \ch{CO} slab model is shown in orange. The J1605 spectra is offset by 8\,mJy for clarity. The slab models are arbitrarily chosen for a visual demonstration of the detections. The vertical dotted lines and the corresponding text indicate the ro-vibrational line transition and the spectral location. The detection of \ch{CO} in J1605 has been previously reported by \citet{2023NatAs...7..805T} and Kanwar et al. (subm.).}
    \label{fig:coemission}
\end{figure}

Figures \ref{fig:hiddenwater_abs} and \ref{fig:coabs} show the stellar photospheric absorption features of \ch{H_2O}, and \ch{CO} in the sample. The models are chosen to match the absorption features visually, but detailed stellar modeling is required to determine whether there can be any \ch{CO} and \ch{H_2O} emission from the disk. Fig.\,\ref{fig:coemission} shows the disk emission of the $P-$branch lines of \ch{CO} ro-vibrational band in J1605 and J1558 \hl{(see Morales-Calder\'on et al. in prep. for analysis of the non-detection of \ch{CO} in NC1)}. \citet{2023NatAs...7..805T} and Kanwar et al. (subm.) report detections of CO $\nu$=1-0 and $\nu$=2-1 lines in J1605. In J1558, the CO $\nu$=2-1 lines are not as prominent as in J1605, but the $\nu$=1-0 lines can be observed from P25 up to P51.

\FloatBarrier
\section{Longer wavelengths: OH and water emission}
\label{app:waterlongwave}
\begin{figure}[!ht]
    \centering
    \includegraphics[trim={0 0.2cm 0 0},clip,width=\linewidth]{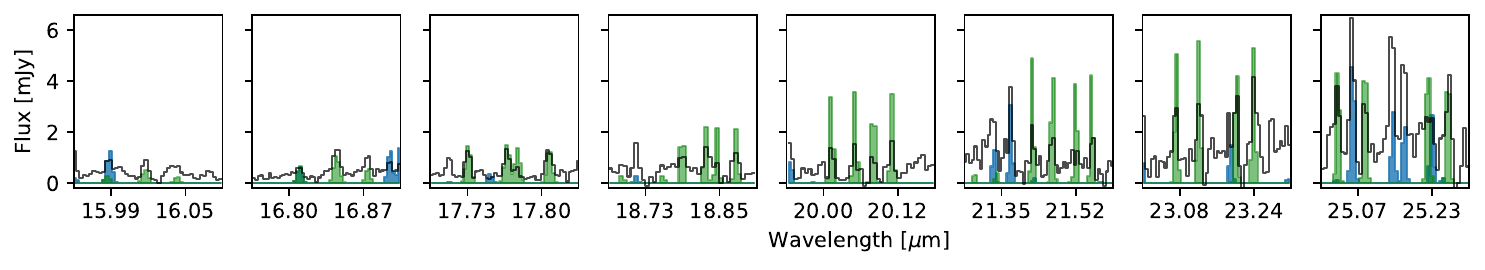}
    \caption{Detection of \ch{OH} lines in J1558. The OH slab model is shown in green (10$^{17}$\,cm$^{-2}$, 800\,K) and the continuum subtracted spectra in black. The water slab model is also shown in each panel (blue).}
    \label{fig:ohemission}
\end{figure}

\begin{figure}[!ht]
    \centering
    \includegraphics[width=0.4\linewidth]{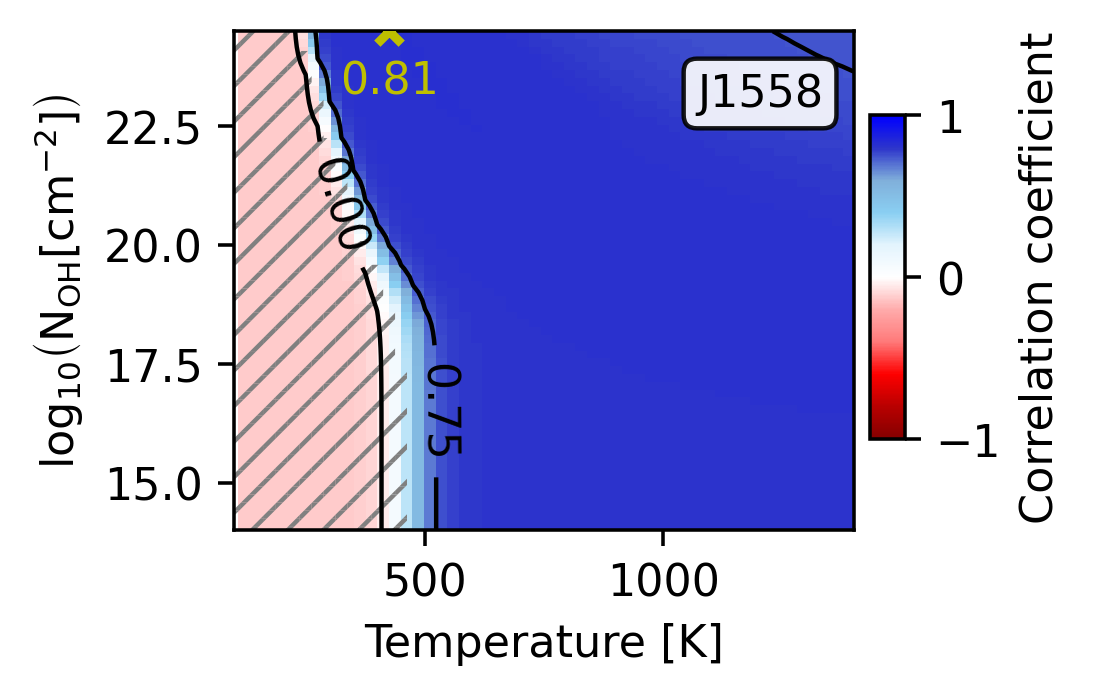}
    \caption{\hl{Cross-correlations of J1558 spectra with OH slab models in the wavelength regions shown in Fig.\,\ref{fig:ohemission} for different column densities ($N$) and temperatures ($T$). The correlation coefficient varies from -1 (anti-correlation, in red) to +1 (correlation, in blue). The contours of correlation coefficients at 0, and 0.75 are shown with dark gray lines. The parameter space where the statistical significance is low (i.e., the p-value is larger than 0.03) is hatched. The yellow cross indicates the best correlation coefficient.}}
    \label{fig:ohcorr}
\end{figure}

Figure\,\ref{fig:ohemission} shows the detections of \ch{OH} pure rotational lines from $\sim$16-25.5\,$\mu$m in the continuum subtracted spectrum of J1558. \hl{We find clear match between several quadruple line pairs of the \ch{OH} slab model and the observed spectra. We also find statistically significant correlation coefficients larger than 0.75 between \ch{OH} slab models and the observed spectrum for a wide range of combinations of temperatures and column densities (Fig.\,\ref{fig:ohcorr}). The OH LTE slab models do not reproduce the observed fluxes of some lines, likely due to significant non-LTE excitation effects \citep{2021A&A...650A.192T, 2024A&A...691A..11T}, which our slab models do not account for. However, these effects have minimal impact on cross-correlations.} 

\begin{figure}[!ht]
    \centering
    \includegraphics[width=\linewidth]{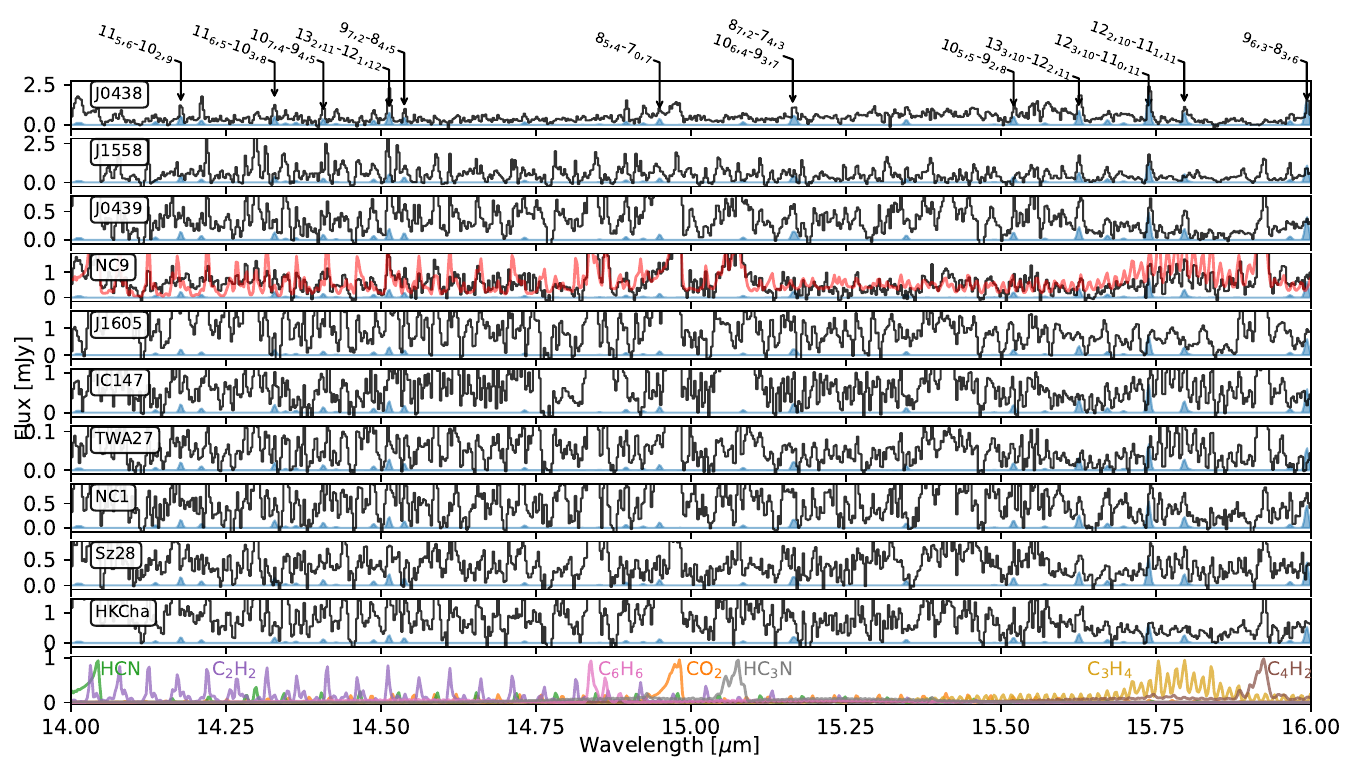}
    \caption{Same as Fig.\,\ref{fig:hiddenwater}, but in the wavelength range 14-16\,$\mu$m. \hl{The bottom panel shows synthetic spectra of molecules detected in this wavelength range, each normalized to their peaks. The sum of these spectra is shown in the fourth panel (red).}}
    \label{fig:hiddenwater2}
\end{figure}

\begin{figure}[!ht]
    \centering
    \includegraphics[width=\linewidth]{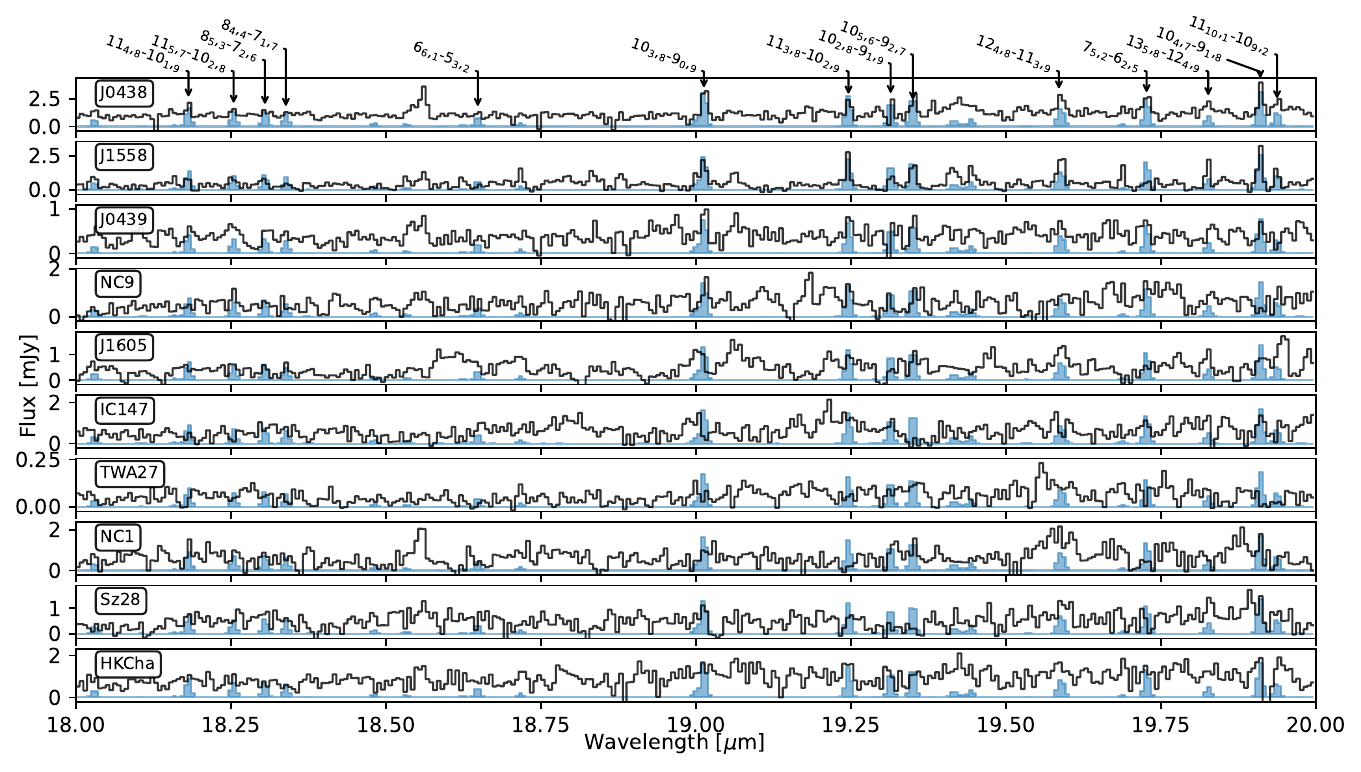}
    \caption{Same as Fig.\,\ref{fig:hiddenwater}, but in the wavelength range 18-20\,$\mu$m.}
    \label{fig:hiddenwater20}
\end{figure}

\begin{figure}[!h]
    \centering
    \includegraphics[width=\linewidth]{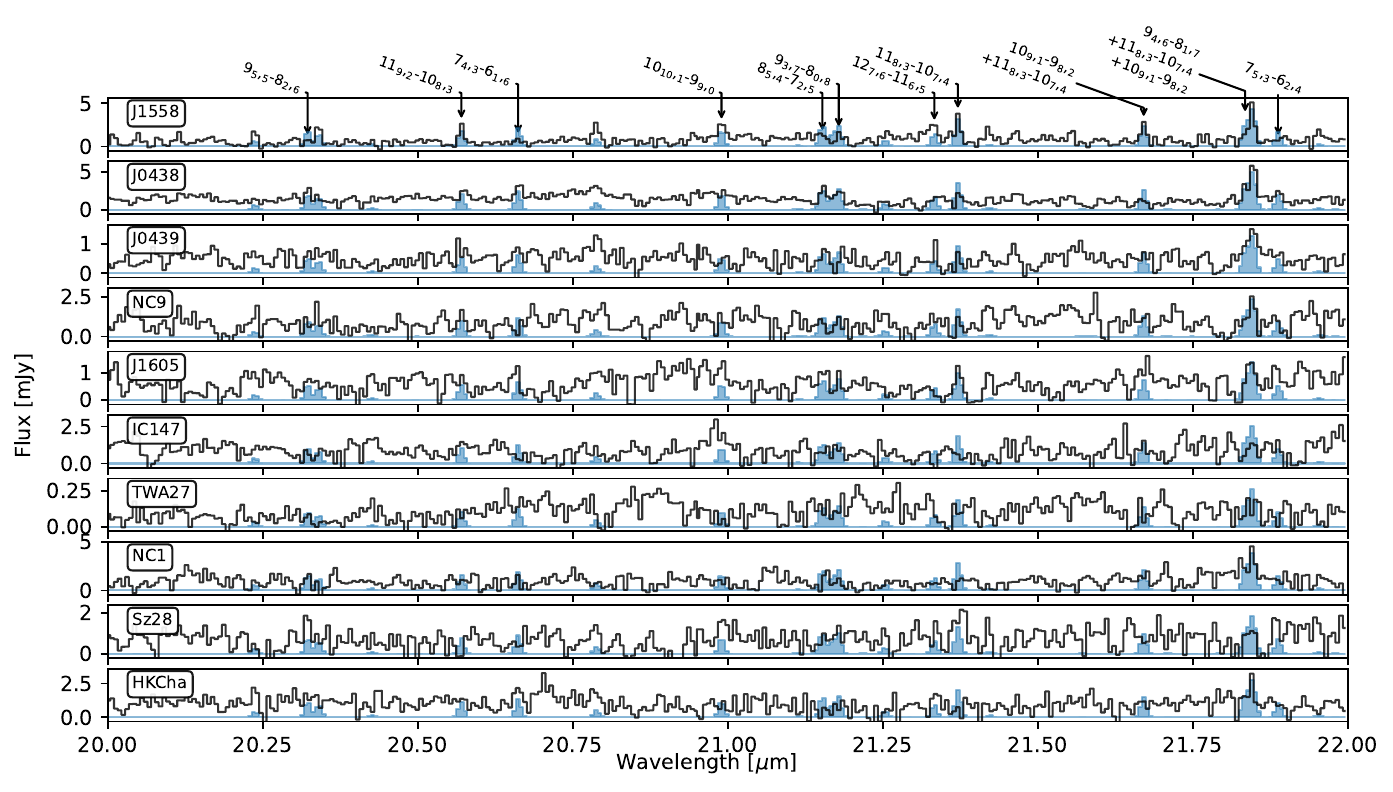}
    \caption{Same as Fig.\,\ref{fig:hiddenwater}, but in the wavelength range 20-22\,$\mu$m.}
    \label{fig:hiddenwater22}
\end{figure}

\begin{figure}[!ht]
    \centering
    \includegraphics[width=\linewidth]{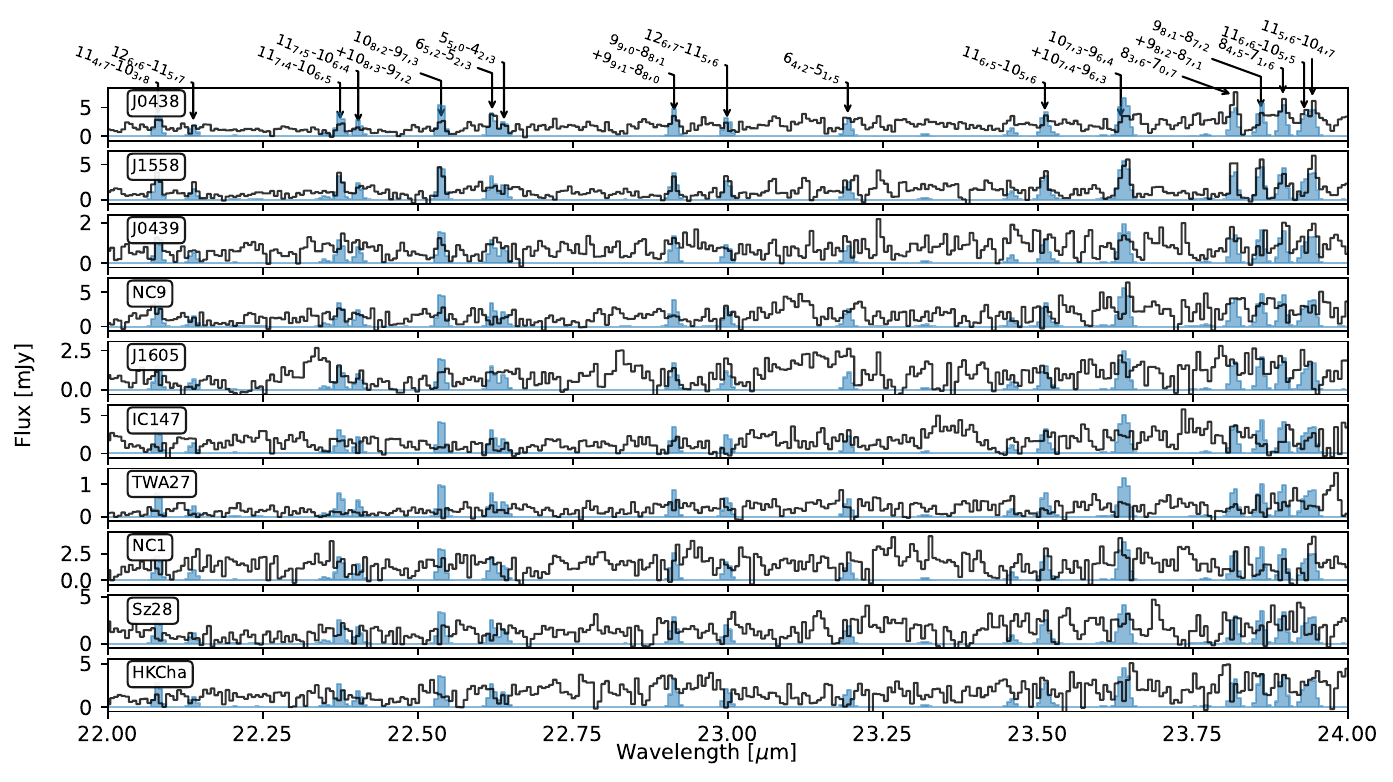}
    \caption{Same as Fig.\,\ref{fig:hiddenwater}, but in the wavelength range 22-24\,$\mu$m.}
    \label{fig:hiddenwater24}
\end{figure}

\begin{figure}[!h]
    \centering
    \includegraphics[width=\linewidth]{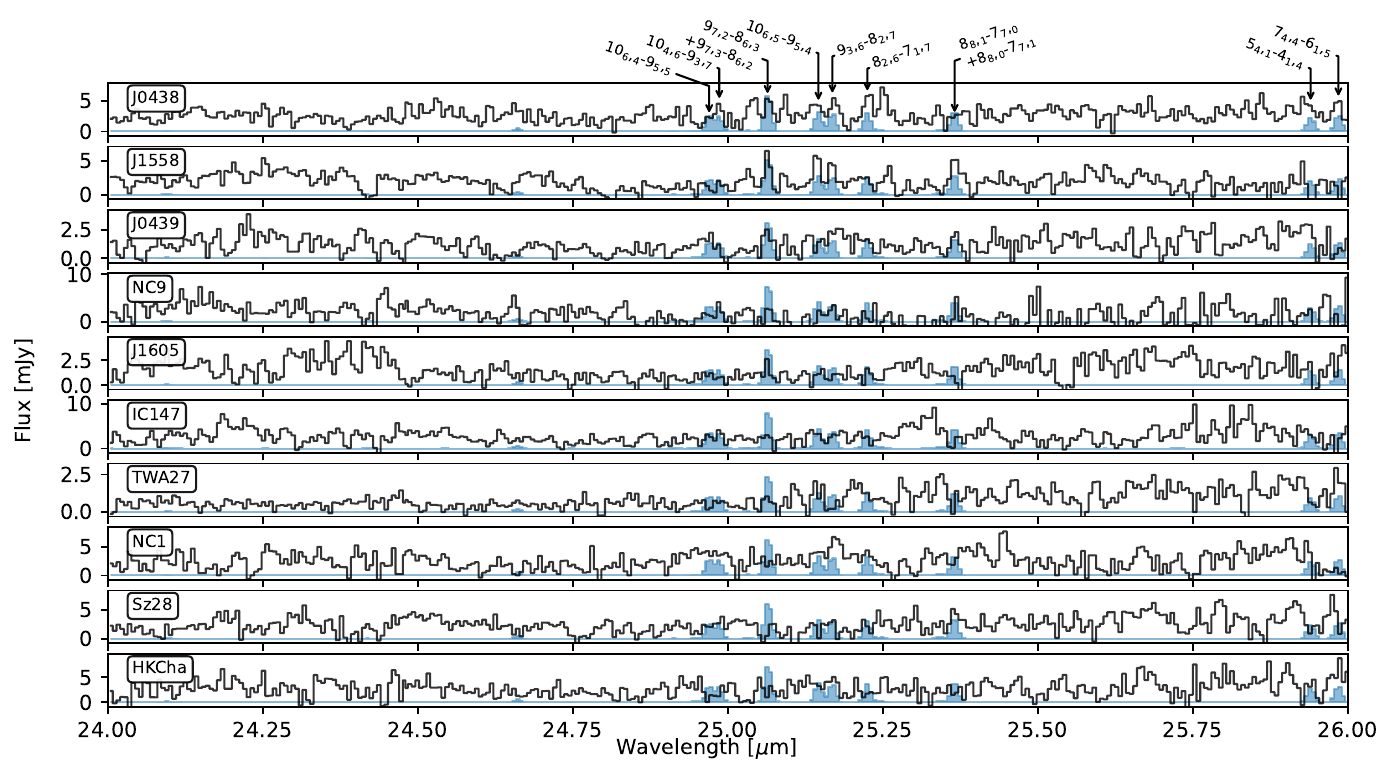}
    \caption{Same as Fig.\,\ref{fig:hiddenwater}, but in the wavelength range 24-26\,$\mu$m.}
    \label{fig:hiddenwater26}
\end{figure}

Figs.\,\ref{fig:hiddenwater2}-\ref{fig:hiddenwater26} show the continuum subtracted spectra along with the water slab models between 14-16\,$\mu$m, and 18-26\,$\mu$m. In Fig.\,\ref{fig:hiddenwater2}, the top panel shows the spectrum of J0438, which has the weakest \ch{C_2H_2} feature and no other hydrocarbon detection. Water lines are detectable in this spectrum at all wavelengths presented in the figure. In the second panel, J1558, the source has a bright \ch{C_2H_2} feature but only one other hydrocarbon, and a weak \ch{CO_2} feature. In this case, the water lines are observed beyond $\sim$15.5\,$\mu$m. In the rest of the sources, the hydrocarbons and \ch{CO_2} dominate, and thus almost no water lines are detected in this range. \hl{The fourth panel shows the spectrum of NC9, in which water is tentatively detected. We also show there the sum of synthetic spectra (at 10$^{17.5}$\,cm$^{-2}$ and 325\,K) of the molecules detected in the NC9 spectrum in the wavelength range 14-16\,$\mu$m - \ch{C_2H_2}, \ch{HCN}, \ch{C_6H_6}, \ch{CO_2}, \ch{HC_3N}, \ch{C_3H_4}, and \ch{C_4H_2} (see Paper 1) - each normalized to its peak. The synthetic spectrum clearly illustrates how the carbon-bearing species can hide water emission.} Beyond 18\,$\mu$m (Figs.\,\ref{fig:hiddenwater20}-\ref{fig:hiddenwater26}), several features match between the observations and the slab models; however, the noise levels increase in most of the sources, and reliable water line detections cannot be made. Reliable water detections can be made in J1558 up to $\sim$25\,$\mu$m.

\begin{figure}[!h]
    \centering
    \includegraphics[width=\linewidth]{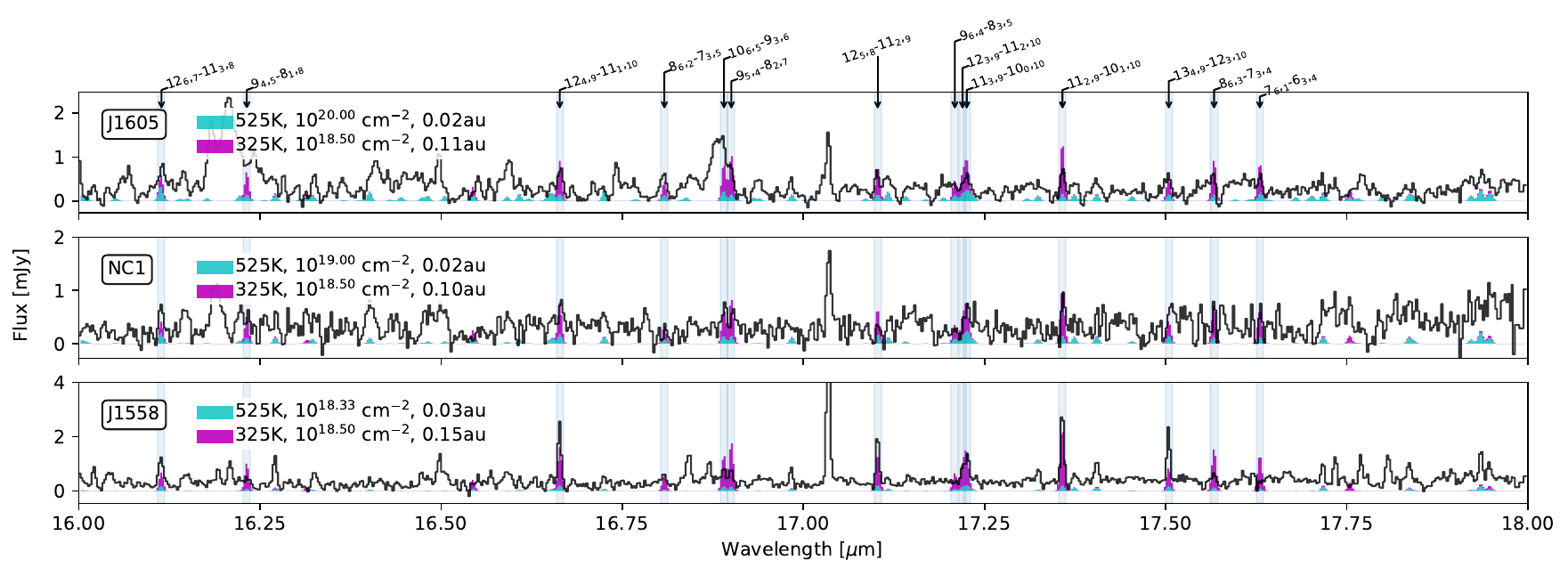}
    \caption{Thermal gradient in VLMS water emission. The hot water spectra (cyan) are same as Fig.\,\ref{fig:hiddenwater_rovib} obtained by visual fit to the ro-vibrational band and the cold water spectra (magenta) are same as in Fig.\,\ref{fig:hiddenwater} but scaled to match the flux deficit between the observed spectra (black) and hot water components. The line positions of pure rotational lines detected in Fig.\,\ref{fig:hiddenwater} are also highlighted.}
    \label{fig:hiddenwater3}
\end{figure}

Figure\,\ref{fig:hiddenwater3} shows the contribution of the higher temperature (525\,K) slab models presented in Fig.\,\ref{fig:hiddenwater_rovib} to the longer wavelengths where the pure rotational lines are detected or are at least expected to be emitted (sources J1558, J1605, and NC1). These higher temperature models do not reproduce the spectral features at these longer wavelengths. \hl{Increasing the emitting area of these models would increase the ro-vibrational strengths. Consequently, we need additional slab components with larger emitting areas at lower temperatures to match the fluxes at those wavelengths, indicating possible thermal gradients in water emission.}

\FloatBarrier

\section{Correlation with water slab models}
\label{app:corr}
\subsection{A check for false positives}
\begin{figure}[htbp]
    \centering
    \includegraphics[width=0.99\linewidth]{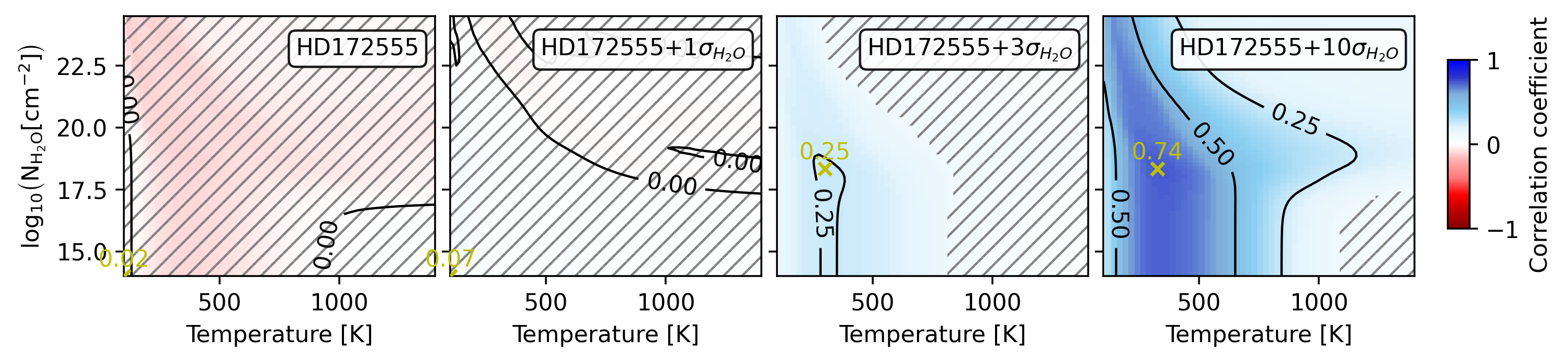}
    \caption{Cross-correlation of the observed spectrum of HD172555 (Samland et al. in subm.) with water slab models following the same procedure as in Fig.\,\ref{fig:water_corr}. The panel on the left shows the cross-correlation with the debris disk spectrum and the water slab spectra, while in the remaining three panels we have added a synthetic water slab spectra at 325\,K and column density of $10^{18.5}$\,cm$^{-2}$ and varying strengths. \hl{Here 1$\times$, 3$\times$, and 10$\times$ $\sigma_{H_2O}$ represent the synthetically added water spectra, each normalized to 1, 3, and 10 times the noise level, respectively}.}
    \label{fig:hd172555}
\end{figure}

In Sect.\,\ref{sec:crosscorr}, we investigate the correlation coefficients between the observed spectra and a grid of water slab model spectra. Here we check for false positives due to noise and other molecular features. For the purpose of noise, we perform the cross-correlations between the debris disk spectra of HD172555 (\hl{devoid of molecular emission,} Samland et al. subm.) and water slab spectra \hl{in the same wavelength range as in Sect.\,\ref{sec:crosscorr}}, shown in Fig.\,\ref{fig:hd172555}. \hl{We added a synthetic water slab with varying strengths to test the sensitivity of the correlation coefficients to the water strengths and find that} water at 3$\sigma$ produces a statistically significant correlation coefficient $\sim$0.25 \hl{(see panel 3 of Fig.\,\ref{fig:hd172555})}. We also performed the same cross-correlation between the debris disk and water slab spectra by removing the two emission lines in the 16.65-17.65\,$\mu$m range, which enhances the noise level when normalized. The cross-correlation remained mostly the same.

\begin{figure}[htbp]
    \centering
    \includegraphics[width=0.8\linewidth]{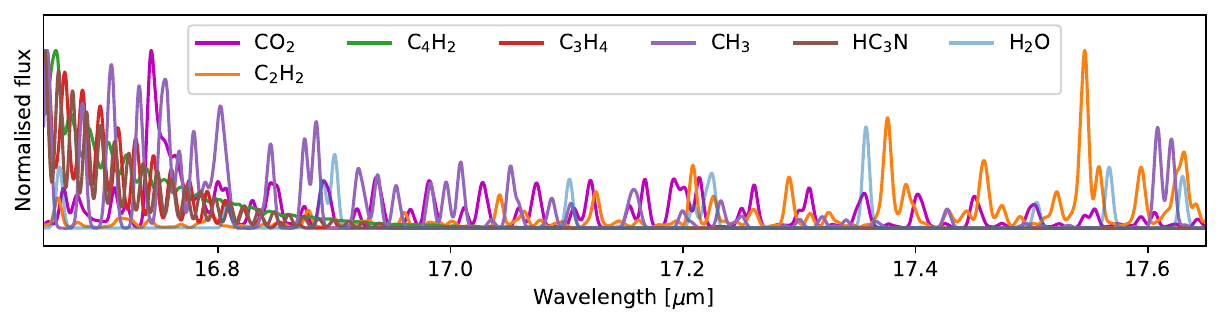}
    \caption{The individual synthetic spectra of carbon-bearing molecules that is combined to perform the cross-correlation with water slab spectra in Fig.\,\ref{fig:HCcorr}. \hl{For reference, a synthetic water spectrum at 1$\sigma$ level, corresponding to the second panel of Fig.\,\ref{fig:HCcorr}, is shown in blue.}}
    \label{fig:synth_hcs}
\end{figure}

\begin{figure}[htbp]
    \centering
    \includegraphics[width=0.99\linewidth]{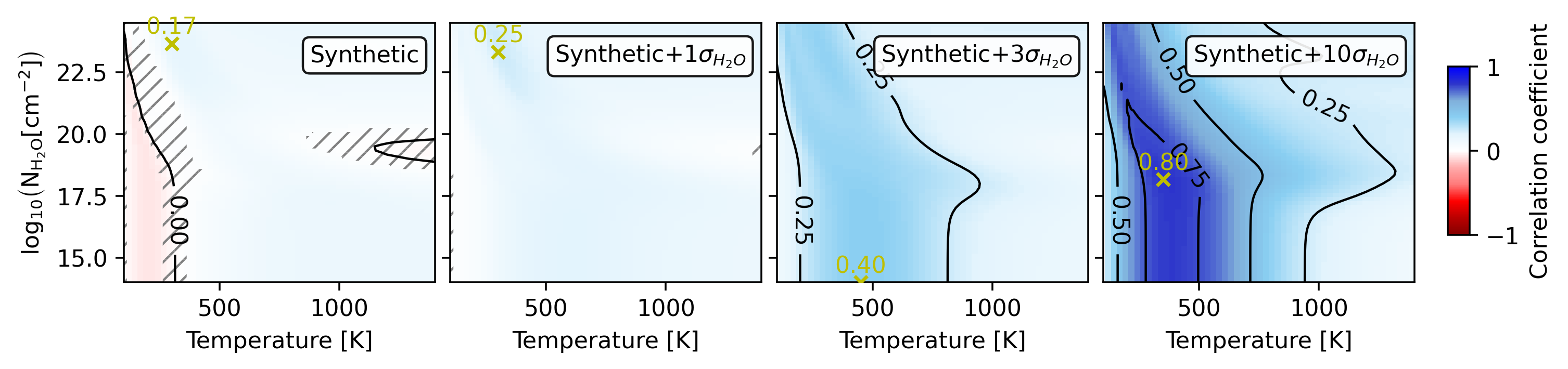}
    \caption{Cross-correlation between synthetic spectra of carbon-bearing molecules with water slab models following the same procedure as in Fig.\,\ref{fig:water_corr}. Similar to Fig.\,\ref{fig:hd172555}, the panel on the left shows the cross-correlation with the synthetic spectrum and the water slab spectra, while in the remaining three panels we have added a synthetic water slab spectra at 325\,K and column density of $10^{18.5}$\,cm$^{-2}$ and varying strengths.}
    \label{fig:HCcorr}
\end{figure}

To test whether false positives could arise because of other molecular emission instead of water, we performed the same tests with synthetic spectra (Fig.\,\ref{fig:synth_hcs}) of the carbon-bearing molecules observed in the sample. Fig.\,\ref{fig:HCcorr} shows the correlation coefficients. We find statistically significant false positive correlation coefficients $<$0.2 due to the emission of these other molecules. However, even including water at 1$\sigma$ level increases the correlation coefficient beyond 0.25. 

Combining both noise (HD172555 spectrum) and molecular emission of other species leads to a decrease in the correlation coefficient and becomes statistically insignificant (p-value$>$0.03). Therefore, we assume that a statistically significant correlation coefficient $\geq$0.25 is an indication of an underlying water emission.

\FloatBarrier

\subsection{Correlation coefficients for ro-vibrational water emission}
\begin{figure}[htbp]
    \centering
    \includegraphics[width=0.99\linewidth]{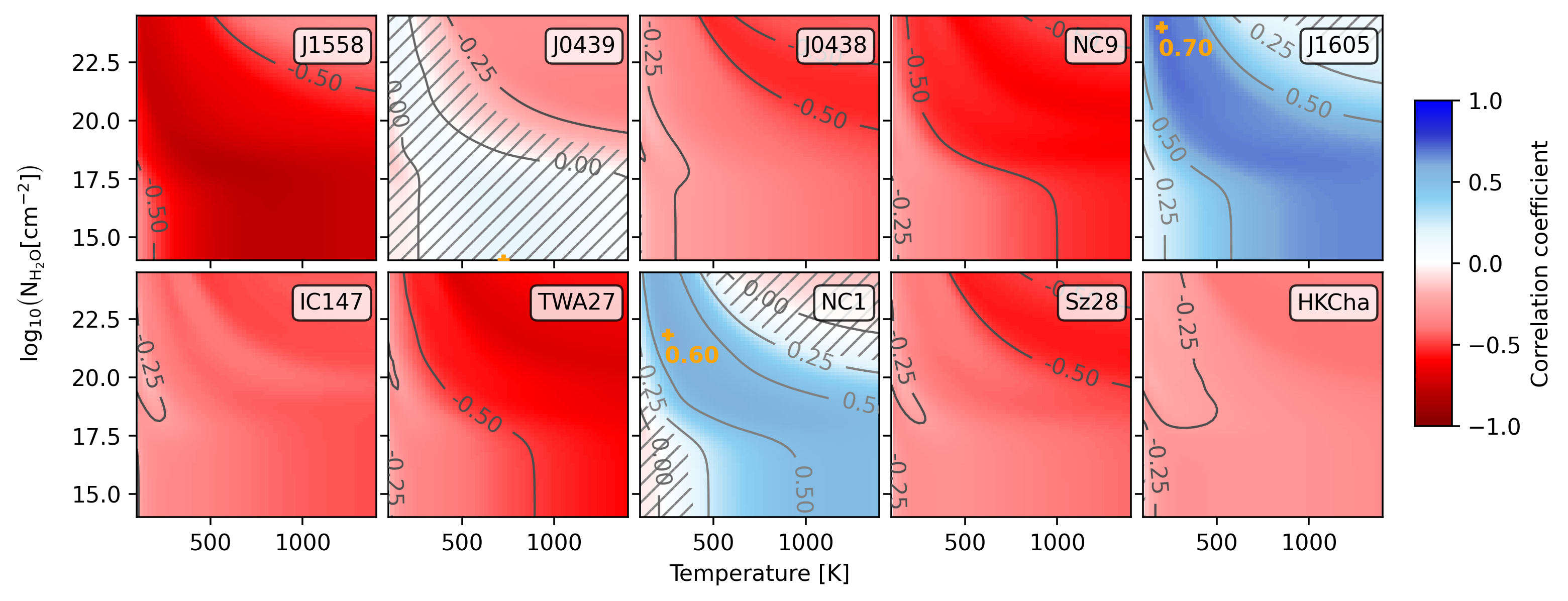}
    \caption{Same as Fig.\,\ref{fig:water_corr} but for the ro-vibrational band emission between 6.55 and 6.70\,$\mu$m. The contours of correlation coefficients at -0.5, -0.25, 0, 0.25, 0.5, and 0.75 are shown with dark gray lines. The orange cross indicates the best correlation coefficient for sources with a positive correlation.}
    \label{fig:Rovibcorr}
\end{figure}

\hl{Figure\,\ref{fig:Rovibcorr} shows the correlation coefficients between the normalized, continuum-subtracted VLMS spectra and the normalized water slab model spectra from 6.55 to 6.70,$\mu$m, similar to Fig.\,\ref{fig:water_corr}. Sources NC1 and J1605 display strong, statistically significant correlations, while sources with stellar absorption features (Fig.\,\ref{fig:hiddenwater_abs}) show strong anti-correlations. In J1558, although part of the parameter space shows positive correlations, they are not statistically significant. However, another region shows a statistically significant anti-correlation. While ro-vibrational water emission in J1558 is a tentative detection based on our criteria, more detailed modeling of ro-vibrational water disk emission and stellar photospheric absorption is needed to fully understand these features.}

\FloatBarrier

\section{\ch{H_2O}/\ch{C_2H_2} flux ratios with slab models}
\label{app:fluxratios}

\begin{figure}[htbp]
    \centering
    \includegraphics[width=0.99\linewidth]{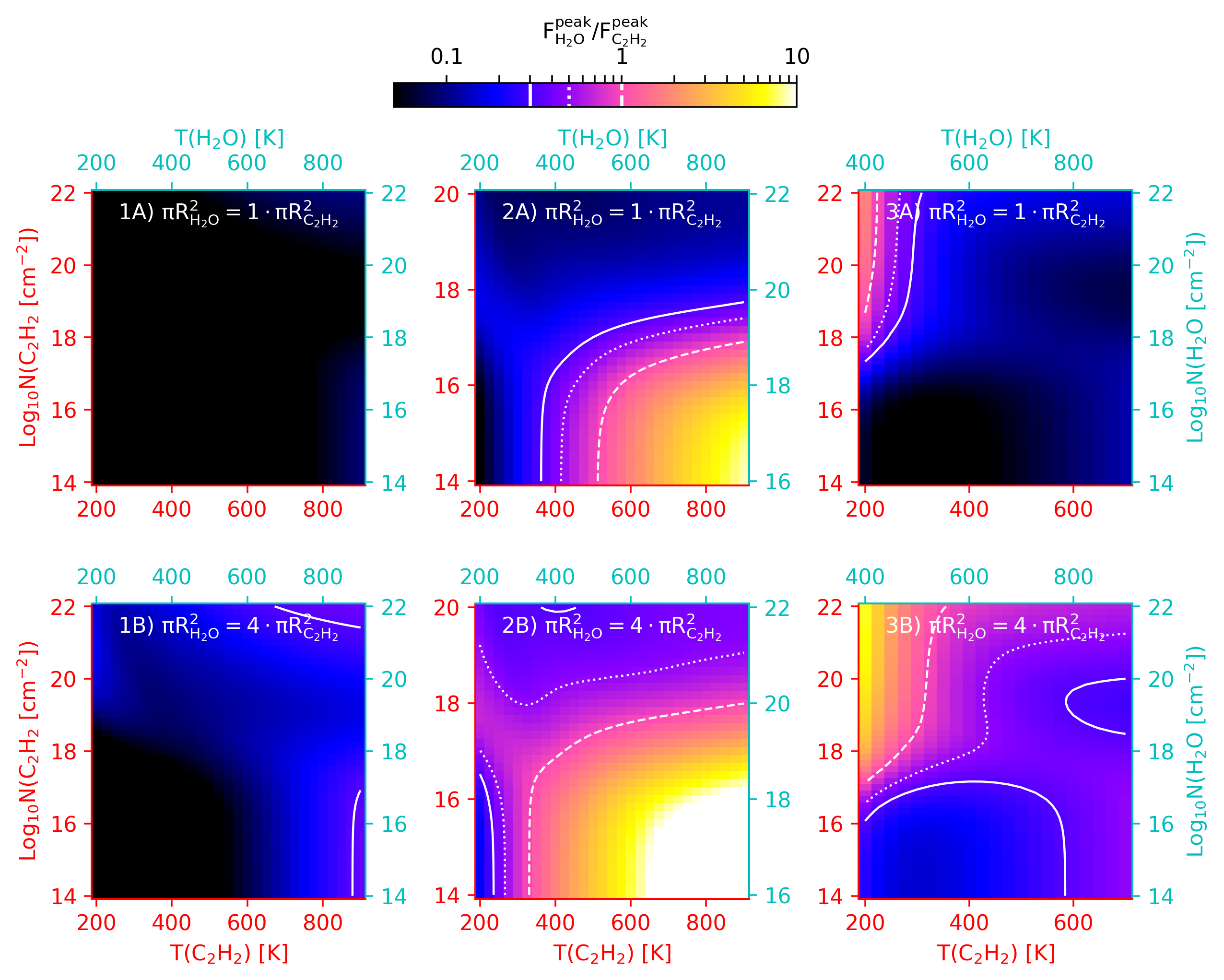}
    \caption{Flux ratios of the peaks of \ch{H_2O} in the 16-18\,$\mu$m wavelengths window and \ch{C_2H_2} in the 13-14\,$\mu$m wavelengths window. The top panels (A) show the peak-flux ratios assuming the same emitting area for both molecules, the bottom panels (B) assume water emitting area four times that of \ch{C_2H_2}. Panels 1A and 1B assume the same column densities and temperatures for both molecules, panels 2A and 2B assume water column densities two orders of magnitude larger than \ch{C_2H_2}, panels 3A and 3B assume water model temperatures higher than the \ch{C_2H_2} by 200\,K. The white contours correspond to flux ratio of 0.3 (solid), 0.5 (dotted) and 1 (dashed).}
    \label{fig:compare1}
\end{figure}

In Sect.\,\ref{sec:slabpeak}, we showed that the water line fluxes could be outshone by the $Q-$ and $P-$branch lines of molecules that have bright emission from their ro-vibrational bands, e.g., \ch{C_2H_2} and \ch{CO_2}. To further investigate how the relative peak fluxes of water and \ch{C_2H_2} behave at different temperatures, column densities, and emitting area, we present in Fig.\,\ref{fig:compare1} the peak flux ratios for three scenarios using two assumptions for the relative size of the emitting area of water and \ch{C_2H_2}: 1) the same emitting areas of water and \ch{C_2H_2}, 2) emitting area of water that is four times that of \ch{C_2H_2}. \hl{The emitting area of water reported in the literature, based on MIRI observations of T Tauri disks and one VLMS disk, are roughly between 0.5 to 10 times the \ch{C_2H_2} emitting area (T Tauri disks: \citealt{Grant2023} - 9$\times$, \citealt{Temmink2024}, \citealt{2024A&A...689A.330T} - 0.7$\times$, \citealt{2024ApJ...977..173C} - 2.5$\times$; VLMS disk: \citealt{2023ApJ...959L..25X} - 1.4$\times$)}. The peak flux is measured between 16 and 18\,$\mu$m for water and at the $Q-$branch for \ch{C_2H_2}. For simplicity, we assume that if the peak water line flux amounts to 30\% of the peak \ch{C_2H_2} flux, it would make it `detectable' in the forest of hydrocarbon lines. Further, we assume water and \ch{C_2H_2} to be at local thermodynamic equilibrium.

\textbf{Case 1: The temperatures and column densities of $\mathbf{C_2H_2}$ and $\mathbf{H_2O}$ are the same.} Panels 1A and 1B of Fig.~\ref{fig:compare1} show the ratio $F_{\rm H_2O}^{\rm peak}/F_{\rm C_2H_2}^{\rm peak}$ for this case. At similar emitting conditions, \ch{C_2H_2} always outshines the pure rotational water lines. Even when the emitting area of water is four times that of \ch{C_2H_2}, the peak fluxes of water lines are less than 30\% of that of \ch{C_2H_2}.

\textbf{Case 2: The column density of water is two orders of magnitude larger than that of $\mathbf{C_2H_2}$.} Panel 2A of Fig.~\ref{fig:compare1} shows that for the same emitting area, \ch{C_2H_2} will outshine water at all temperatures if the \ch{C_2H_2} column density is larger than $\sim$$10^{18}\,\rm cm^{-2}$. The same is true at all column densities if the temperatures are below $\sim$350\,K. In other words, water detection can only be possible at high temperatures ($>$350\,K) and low \ch{C_2H_2} column densities ($<$$10^{18}$\,cm$^{-2}$). However, if the water emitting area is increased by a factor of four (panel 2B) the water peak flux would be at a detectable level across almost the entire parameter space we explored.

\textbf{Case 3: The temperature of water is 200\,K higher than that of $\mathbf{C_2H_2}$, while the column densities are the same.} Panel 3A of Fig.~\ref{fig:compare1} shows that with the same emitting areas, water would be outshone by \ch{C_2H_2} across almost the entire explored parameter space. Only when the column densities are larger than $\rm\sim$$10^{17}$cm$^{-2}$ and the temperature of \ch{C_2H_2} is $\lessapprox$300\,K, water has a detectable flux level. Considering a water emitting area four times that of \ch{C_2H_2}, the water flux becomes stronger and thus detectable in a larger region of the explored parameter space (panel 3B).

The main takeaway from this analysis is that compared to \ch{C_2H_2}, the \ch{H_2O} line fluxes are more sensitive to increasing the emitting area than to increasing the temperature or column densities. This is because the emitting area directly scales the flux, while the column density scales the optical depth of the lines. In case of \ch{H_2O}, the pure rotation lines are spectrally well separated, due to which the line fluxes are more sensitive to the emitting area. In the case of \ch{C_2H_2}, the numerous ro-vibration emission lines and the weaker forest of lines (e.g. hot bands) overlap, making it sensitive to the column density and temperature, which in turn lead to higher flux levels.

\FloatBarrier


\bibliography{sample631}{}
\bibliographystyle{aasjournal}



\end{document}